\documentclass[letterpaper]{article}

\usepackage{titling}
\usepackage{fullpage,authblk}
\usepackage{amssymb,amsfonts,amsmath,enumerate}
\usepackage{graphicx,bm,latexsym}
\usepackage{multirow}
\usepackage{algorithm}
\usepackage{algpseudocode}
\usepackage{tocloft}
\usepackage[bookmarks, colorlinks=true, plainpages = false, citecolor = green, urlcolor = black, filecolor = black]{hyperref}

\DeclareMathOperator*{\argmin}{arg\,min}
\DeclareMathOperator*{\argmax}{arg\,max}

\begin{document}

\title{How Entropic Regression Beats the Outliers Problem \\in Nonlinear System Identification}

\author[1,2]{Abd AlRahman R. AlMomani}
\author[3]{Jie Sun}
\author[1,2]{Erik Bollt}

\affil[1]{Clarkson Center for Complex Systems Science ($C^3S^2$), Potsdam, NY, 13699, USA.}
\affil[2]{Electrical and Computer Engineering, Clarkson University, Potsdam, NY, 13699, USA.}
\affil[3]{Theory Lab, Hong Kong Research Centre of Huawei Tech, Hong Kong, 852, China.}

\date{}

\maketitle

\begin{abstract}
In this work, we developed a nonlinear System Identification (SID) method that we called Entropic Regression. Our method adopts an information-theoretic measure for the data-driven discovery of the underlying dynamics. Our method shows robustness toward noise and outliers and it outperforms many of the current state-of-the-art methods. Moreover, the method of Entropic Regression overcomes many of the major limitations of the current methods such as sloppy parameters, diverse scale, and SID in high dimensional systems such as complex networks. The use of information-theoretic measures in entropic regression poses unique advantages, due to the Asymptotic Equipartition Property (AEP) of probability distributions, that outliers and other low-occurrence events are conveniently and intrinsically de-emphasized as not-typical, by definition. We provide a numerical comparison with the current state-of-the-art methods in sparse regression, and we apply the methods to different chaotic systems such as the Lorenz System, the Kuramoto-Sivashinsky equations, and  the Double Well Potential.\\

\noindent \textbf{Keywords:} System Identification, Sparse Regression, Data-Driven Modeling, Entropy, Conditional Mutual Information, Asymptotic Equipartition Property, Nonlinear dynamics, Complex Systems.
\end{abstract}

%
%
{\bf System identification (SID) is a central concept in science and engineering applications whereby a general model form is assumed, but active terms and parameters must be inferred from observations.
{Most} methods for SID rely on optimizing some metric-based cost function that describes how a model fits observational data. {A commonly used} cost function employs a Euclidean metric and leads to a least squares estimate, whereas recently it has become popular to also account for model sparsity such as in compressed sensing and Lasso. While the effectiveness of these methods has been demonstrated in {previous studies including in cases where outliers exist in sparse samples}, SID remains particularly difficult under more realistic scenarios where {each observation is subject to non-negligible noise, and sometimes even contaminated by} large noise outliers. Here we report that {existing} sparsity-focused methods such as compressive sensing, when {applied in such scenarios,} can result in ``over sparse" solutions that are brittle to outliers. In fact, metric-based methods are prone to outliers because outliers by nature have an unproportionally large influence. To mitigate such issues of large noise and outliers, we develop an entropic regression approach for nonlinear SID, whereby true model structures are identified based on an information theoretic criterion describing relevance in terms of  reducing information flow uncertainty, versus not necessarily (just) sparsity. The use of information-theoretic measures in entropic regression poses unique advantages, due to the asymptotic equipartition property of probability distributions, that outliers and other low-occurrence events are conveniently and intrinsically de-emphasized as not-typical, by definition.}

A basic and fundamental problem in science and engineering is to collect data as observations from an experiment, and then to attempt to explain the experiment by summarizing data in terms of a model.  
\textcolor{black}{When dealing with a dynamical process, a} common scenario is to describe the underlying process as a dynamical system, which may be in the form of a differential equation (DE).  Traditionally this means ``understanding the underlying physics," in a manner that allows one to write a DE from first principles, including those terms to \textcolor{black}{capture} the delicate but important (physical) effects.  Validation of the model may come from comparing outputs from the model to those from experiments, where outputs are typically represented as multivariate time-series. Building a DE model based on fundamental laws and principles requires strong assumptions, which might be evaluated by how the model fits data. Weigenbend and Gershenfeld made a distinction between weak modeling (data rich and theory poor) and strong modeling (data poor and theory rich), and suggest that it is related to ``...the distinction between memorization and generalization..."~\cite{Gershenfeld1993}.

The problem of learning a (dynamical) system from observational data is known as {\it system identification} (SID), and often times involves the underlying assumption that the {\it structural} form of the DE is known (which kinds of terms to include in the functional description of the equation), but only the underlying parameters are not known. For example, suppose we observe the dynamics of a simple dissipative linear spring, then we may express the model as $m \ddot{x}+\gamma \dot{x}+ k x=0$ based on Hooke's law. However, the parameters $m, \gamma,$ and $k$ might be unknown and need to be estimated in order to completely specify the model for purposes such as prediction and control.  One may directly measure those parameters by static testing (e.g., weighing the mass on a scale). Alternatively, here we are interested in utilizing the observational data generated by the system without having to design and perform additional experiments, to estimate the parameters corresponding to the model that best fits empirical observations, which is a standard viewpoint in SID. 
In this thought experiment, the SID process is performed with the underlying physics understood (the form of the Hooke spring equation). In general it can be applied in the scenario where very little information is previously known about the system, in a black box manner.

Suppose that observations $\{\bm{z}(t)\}$ come from a general (multidimensional, coupled) DE, represented by
\begin{equation}\label{eq:main}
\dot{\bm{z}}=\bm{F}(\bm{z}),
\end{equation}
where $\bm{z} = [z_1,\dots, z_N]^{T} \in \mathbb{R}^N$ \textcolor{black}{is the (multivariate) state variable of the system} and $\bm{F}=[F_1,\dots,F_N]^\top:\mathbb{R}^N\rightarrow\mathbb{R}^N$ \textcolor{black}{is the vector field}. Each component function $F_i(\bm{z})$ can be represented using a series expansion (for example a power series or a Fourier series), writing generally,
\begin{equation}\label{eq:basis}
\dot{z}_i=F_i(\bm{z}) = \sum_{k=0}^{\infty}a_{ik}\phi_k(\bm{z}),
\end{equation}
for a linear combination of basis functions $\{\phi_k\}_{k=0}^\infty$.
The basis functions do not need to be mutually orthogonal, \textcolor{black}{and the series can even include multiple bases, for example to contain both a polynomial basis and a Fourier basis~\cite{Brunton2015}}.
The coefficients $\{a_{ik}\}$ are to be determined by contrasting simulations to experimental measurements, in an optimization process whose details of how error is measured distinguishes the various methods we discuss here. 
This was the main theme in previous \textcolor{black}{approaches on nonlinear SID, with different methods differ mainly on how a model's fit is quantified. The different approaches include using standard squared error measures~\cite{BolltYao2007,Chen1989}, sparsity-promoting methods~\cite{Kalouptsidis2011,Brunton2015,Wang2011,Wang2016} as well as using entropy-based cost functions~\cite{guo2008extended}. 
Among those, sparsity-promoting methods have proven particularly useful because they tend to avoid the issue of overfitting, thus allowing a large number of basis functions to be included to capture possibly rich dynamical behavior~\cite{Kalouptsidis2011,Kalouptsidis2011,Brunton2015,Wang2011}.} 

\textcolor{black}{Regardless of the particular method or system, most previous work on nonlinear SID focused on the low-noise regime and demonstrated success only when there is a sufficient amount of clean observational data.} 
In practice, an observation process can be subject to external disturbances in unpredictable ways. Consequently, the effective noise can be quite large and even with frequently occurring ``outliers'' both of which may contaminate the otherwise perfect data. Can SID still work under the presence of large noise and outliers? At a glance, the answer should be yes, given that several recent SID methods for nonlinear systems are readily deployable in the presence of noise. For example, compressive sensing can handle noise by relaxing the constraint set \textcolor{black}{whereas least squares and Lasso can be applied off the shelf---the important question however is whether the quality of solution is compromised or not, and to what extent. Recently Tran and Ward considered the nonlinear SID problem under the presence of outliers in observational data and showed that so long as there the outliers are ``sparse" leaving sufficient amount of ``clean"  data available, existing techniques such as SINDy can be extended to reconstruct the exact form of a system with high probability~\cite{TranWard}. In the current work, we are interested in the more realistic scenario where effective noise is present everywhere and thus {\it all} data points are contaminated by non-negligible noise and sometimes outliers. These features effectively creates a ``high noise and low data amount" regime, where we found that existing nonlinear SID methods including recent ones that specialize in promoting sparsity, fall short.}

In this work we depart from most standard approaches for nonlinear SID. We identify the error quantification via metric-based cost functions as a root cause of existing methods to fail under large noise and outliers because outliers tend to deviate from the rest of sample data as measured by metric distance; thus trying to ``fit'' the outliers \textcolor{black}{almost inevitably} causes the model to put (much) less weights on the ``good'' data points. 
\textcolor{black}{To resolve this important issue}, we propose to infer the (sparsity) structure of a general model together with its parameters using a novel {\it information theoretic regression} approach that we call Entropic Regression (\textcolor{black}{ER}).
As we will show, while standard metric-based methods emphasize the data in ways as designed by the chosen metric, the proposed ER approach is robust with regards to the presence of noise and outliers in the data. Instead of searching for the sparsest model and thus risk forcing a wrong sparse model, ER is emphasizing ``information relevance'' according to a model-free, entropic criterion. Basis terms will be included in the model only because they are relevant and not (necessarily) because they together make up the sparsest model. We demonstrate the effectiveness of ER in several examples, including chaotic Lorenz systems, Kuramoto-Sivashinsky equations, and a double well potential, where in each case the observed data contains relatively large noise and outliers. We also remark on the computational complexity and convergence in small-data regime, as well as \textcolor{black}{discuss open problems and future directions}.
\section*{Results}\label{sec:results}
\subsubsection*{Nonlinear System Identification: Problem Statement and Formulation}
\textcolor{black}{Following the standard routine in nonlinear SID~\cite{ljung1999system},} the starting point is to recast the nonlinear SID problem into a computational inverse problem,
by considering an appropriate set of basis functions that span the space of functions including the system of interest~\cite{BolltYao2007,Wang2016}.
A common choice is the standard {\it polynomial basis}
\begin{equation}\label{eq:polybasis}
\bm{\phi}=[\phi_0(\bm{z}),\phi_1(\bm{z}),\phi_2(\bm{z}),\dots]=[1,z_1,z_2,\dots,z_N,z_1z_2,z_1z_3,\dots,z_{N-1}z_{N},\dots]
\end{equation}
\textcolor{black}{where each term is a monomial.}
Using a set of basis functions, one can represent the individual component functions of $F$ as a series as in~\eqref{eq:basis}. 
The specification of the location of nonzero parameters are referred to as the {\it structure} of the model.

Consider time series data $\{\bm{z}(t)=[z_1(t),\dots,z_m(t)]^\top\}_{t=t_0,\dots,t_\ell}$ and corresponding $\{\bm{F}(\bm{z}(t)\}_{t=t_0,\dots,t_\ell}$ generated from a nonlinear, high-dimensional dynamical system~\eqref{eq:main}, possibly subject to observational noise. From $\bm{z}(t)$, one can estimate the derivatives by any of the standard Newton-Cotes methods, explicit Euler's method of course being the simplest, giving $F_i(\bm{z}(t_k))=\frac{z_i(t_{k+1}) - z_i(t_k)}{\tau_k}+\mathcal{O}((t_{k+1}-t_k))$, \textcolor{black}{or central difference which has improved accuracy: $F_i(\bm{z}(t_k))=\frac{z_i(t_{k+1}) - z_i(t_{k-1})}{t_{k+1}-t_{k-1}}+\mathcal{O}((t_{k+1}-t_{k-1})^2)$.}
The problem of nonlinear system identification is to reconstruct the functional form as well as parameters of the underlying system, that is, to infer the nonlinear function $\bm{F}$.

\begin{figure}[ht]
\centering
\includegraphics[width=0.85\textwidth]{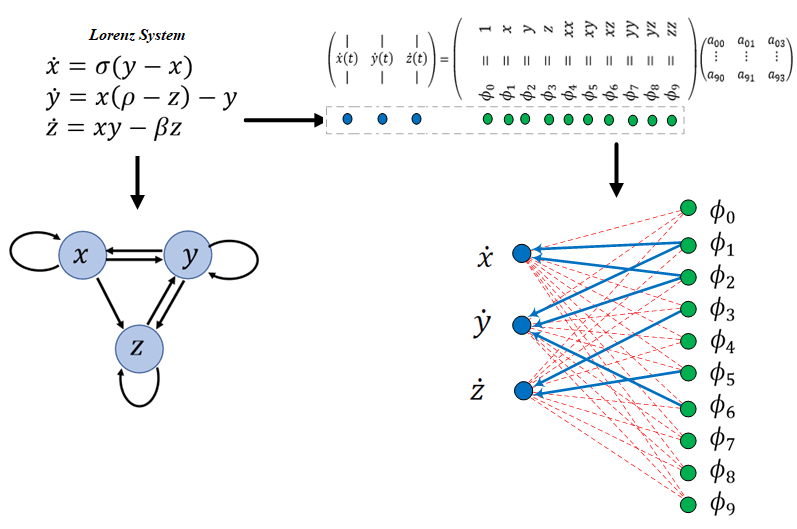}
\caption{(Left) Lorenz system as a dynamical system and its standard graph representation. (Right) Linear combination of nonlinear basis functions, with coupling coefficients \textcolor{black}{$\{a_{ik}\}$ forming the structure of the system (bottom right). Here each directed edges represent the presence of basis terms on the individual variables of the system.}
}
\label{fig:Lorenz}
\end{figure}

Under the basis representation~\eqref{eq:basis}, the identification of $\bm{F}$ becomes equivalent \textcolor{black}{to} estimating all the parameters $\{a_{ik}\}$. 
\textcolor{black}{In practice, the empirically observed state variable is subject to noise: $\hat{\bm{z}}(t)=\bm{z}(t)+\bm{\eta}(t)$ with $\bm{\eta}(t)$ representing the (multivariate) noise and $\hat{F}_i$ denoting the approximated value of $F_i$. For noisy observations $\hat{\bm{z}}(t)$, the difference between $\hat{F}_i(\hat{\bm{z}}(t))$ and $F_i(\hat{\bm{z}}(t))$ originates from several sources: the infinite series is truncated and the derivatives are estimated numerically and by using approximate states. Nevertheless, we can represent the aggregated error as an effective noise $\bm{\xi}(t)$ term and express the forward model as}
\begin{equation}\label{eq:forwardmodel}
\hat{F}_i(\hat{\bm{z}}(t)) = \sum_{k=0}^{K}a_{ik}\phi_k(\hat{\bm{z}}(t)) + \xi_i(t),~(t=t_0,\dots,t_\ell;i=1,\dots,N).
\end{equation}
\textcolor{black}{Note that because of the combined and accumulated effects of observational noise, approximation error and truncation, even if the observational noise of the states $\eta_i(t)$ are iid, this is not necessarily true for the effective noise $\xi_i(t)$.}
\textcolor{black}{In matrix form, the forward model~\eqref{eq:forwardmodel} has the approximate expression}
\begin{eqnarray}
\left(  \begin{array}{ccc}| & | & | \\ \dot{z}_{1}(t_{i}) & \dots & \dot{z}_{N}(t_{i})  \\ | & | & | 
\end{array} \right) &\approx& \left(  \begin{array}{cccc} | &| & | & | \\ \phi_{0}(t_{i}) & \phi_{1}(t_{i}) & \dots & \phi_{K}(t_{i})  \\ | &  | & | & | \end{array} \right) 
\left(  \begin{array}{cccc} a_{00} & a_{01} & \dots & a_{0N} \\ 
\vdots & \vdots & \ddots & \vdots  \\ 
a_{K0} & a_{K1} & \dots & a_{KN} 
\end{array} \right). \notag \\
\end{eqnarray}
Figure~\ref{fig:Lorenz} shows the structure of the Lorenz system under standard polynomial basis up to quadratic terms.

In vector form, under a choice of basis and truncation, the nonlinear system identification problem can be recast into the form of a linear inverse problem
\begin{equation}\label{eq:inverse_i}
\bm{f}^{(i)} = \Phi\bm{a}^{(i)} + \bm\xi^{(i)},
\end{equation}
\textcolor{black}{where $\bm{f}^{(i)}=[\hat{F}_i(\hat{\bm{z}}(t_1)),\dots,\hat{F}_i(\hat{\bm{z}}(t_\ell))]^\top\in\mathbb{R}^{\ell\times1}$ represents the $i$-th component of the estimated vector field from the observational data,
$\Phi=[\bm{\phi}^{(1)},\dots,\bm{\phi}^{(K)}]\in\mathbb{R}^{\ell\times K}$ (with $\bm{\phi}^{(k)}=[\phi_k(\hat{\bm{z}}(t_1)),\dots,\phi_k(\hat{\bm{z}}(t_\ell))]\in\mathbb{R}^{\ell\times1}$) represent sampled data for the basis functions,
$\bm\xi^{(i)}=[\xi_i(t_1),\dots,\xi_i(t_\ell)]^\top\in\mathbb{R}^{\ell\times1}$ represents noise, and
$\bm{a}^{(i)}=[a_{i1},\dots,a_{iK}]^\top\in\mathbb{R}^{K\times1}$ is the vector of parameters which is to be determined. Note that the form of the equation~\eqref{eq:inverse_i} is the same for each $i$, and solving each $\bm{a}^{(i)}$ can be done separately and independently for each $i$. In what follows we omit the index when discussing the general methodology,} and consider the following linear inverse problem
\begin{equation}\label{eq:inverse}
\bm{f} = \Phi\bm{a} + \bm\xi,
\end{equation}
where $\bm{f}\in\mathbb{R}^{\ell\times1}$ and $\Phi\in\mathbb{R}^{\ell\times K}$ are given, with the goal to estimate $\bm{a}\in\mathbb{R}^{K\times1}$.
This general problem is in the form of an inverse problem and is typically solved under various assumptions of noise by methods such as 
least squares, orthogonal least squares, lasso, compressed sensing, to name a few. Each of these methods, \textcolor{black}{in addition to the recent approach of SINDy and its generalization, is} mentioned in the Results section and reviewed in the Methods section. In what follows we develop a unique information-theoretic approach called entropic regression, which we demonstrate has significant advantages.
\subsubsection*{Entropic Regression}
To overcome the competing challenges of potential overfitting, efficiency when limited data points are given, and robustness with respect to noise and in particular outliers in observations, we propose a novel framework that combines the advantage of information-theoretic measures and iterative regression methods. The framework, which we term {\it entropic regression} (ER), is model-free, noise-resilient, and efficient in discovering a ``minimally sufficient" model to represent data. 
\textcolor{black}{The key idea is that, for given set of basis functions, a model should be considered minimally sufficient if no basis function that is not already included in the model can help increase the information relevance between the model outputs and observed data. In other words, the residual between the model fit and observational data is statistically independent from any basis function that is not included in the model---because otherwise the dependence can be harvested to reduce the discrepancy by including such a basis function in the model. We emphasize that, although the idea seems related to classical model selection principles such as AIC~\cite{AIC1974}, ours combines model construction with selection. In addition, even though it is not uncommon for entropy measures to be adopted in system identification~\cite{guo2008extended,prando2017maximum}, the proposed method is unique as it fuses entropy optimization with regression in a principled manner that enables scalable computation and efficient estimation in reconstruction nonlinear dynamics under noisy data. As we shall see below, the proposed ER method is applicable even in the small-sampling regime (by adopting appropriately defined entropy measures and efficient estimators) and naturally allows for a computationally efficient procedure to build up a model from scratch.}
In particular, we use (conditional) mutual information as an information-theoretic criterion and iteratively select relevant basis functions, analogous to the optimal causation entropy algorithm previously developed for causal network inference~\cite{Sun2014b,Sun2014a} \textcolor{black}{but here including an additional regression component in each step.}
Thus, ER can be thought of as an information-theoretic extension of the orthogonal least squares regression, or as a regression version of optimal causation entropy. 

We now present the details of ER. The ER method contains two stages (also see Algorithm~\ref{alg:ER} for the pseudocode): forward ER and backward ER. In both stages, selection and elimination are based on an entropy criterion and parameters are updated in each iteration using a standard regression (e.g., least squares). 
Consider the inverse problem~\eqref{eq:inverse}. 
\textcolor{black}{For an index set $S\subset\mathbb{N}\cup\{0\}$, the estimated parameters can be thought of as a mapping from the joint space of $\Phi$, $\bm{f}$ and $S$ to a vector denoted as $\hat{\bm{a}}=R(\Phi,\bm{f},S)$. For instance, under a least-squares criterion the mapping is given by $R(\Phi,\bm{f},S)_S=\Phi_S^\dagger\bm{f}$ ($\Phi_S$ denotes the columns of matrix $\Phi$ indexed by $S$) and $R(\Phi,\bm{f},S)_i=0$ for all $i\notin S$. Using the estimated parameters, the recovered signal can be computed as $\Phi R(\Phi,\bm{f},S)$.
In the ER algorithm, we start by selecting a basis function $\bm\phi_{k_1}$ that maximizes its mutual information with $\bm{f}$, compute the corresponding parameter $a_{k_1}$ using the least squares method, and obtain the corresponding regression model output $\bm{z}_1$ according to
\begin{equation}
\begin{cases}
k_1 = \argmax_{k}I(\Phi R(\Phi,\bm{f},\{k\});\bm{f}),\\
\hat{\bm{a}} = R(\Phi,\bm{f},k_1),\\
\bm{z}_{1}=\Phi R(\Phi,\bm{f},k_1).
\end{cases}
\end{equation}
\textcolor{black}{Here $I(\bm{x};\bm{y})$ denotes mutual information between $\bm{x}$ and $\bm{y}$, which is a model-free measure of the statistical dependence between two distributions (that is, $\bm{x}$ and $\bm{y}$ are independent if and only if their mutual information equals zero)~\cite{Cover2005}.}
Next, in each iteration of the forward stage, we perform the following computations and updates, for $i=2,3,\dots$,
\begin{equation}
\begin{cases}
k_i = \argmax_{k\notin\{k_1,\dots,k_{i-1}\}}I(\Phi R(\Phi,\bm{f},\{k\}); \bm{f} |\bm{z}_{i-1}),\\
\hat{\bm{a}} = R(\Phi,\bm{f},\{k_1,\dots,k_i\}),\\
\bm{z}_{i}=\Phi R(\Phi,\bm{f},\{k_1,\dots,k_i\})
\end{cases}
\end{equation}
The process terminates when $\max_{k}I(\Phi R(\Phi,\bm{f},k); \bm{f}|\bm{z}_{i-1})\approx 0$ (or when all basis functions are exhausted), indicating that none of the remaining basis function is {\it relevant} given the current model, in an information-theoretic sense.
The result of the forward ER is a set of indices $S=\{k_1,\dots,k_m\}$ together with the corresponding parameters $a_{k_1},\dots,a_{k_m}$ ($a_j=0$ for $j\notin S$) and model $f\approx a_{k_1}\bm\phi_{k_1}+\dots+a_{k_i}\bm\phi_{k_i}.$ Finally, we turn to the backward stage, where the terms that had previously been included are re-examined for their information-theoretic relevance and these that are redundant will be removed. 
In particular, we sequentially check for each $j=k_i\in S$ to determine if the basis term $\bm\phi_j$ is redundant by computing 
\begin{equation}
\begin{cases}
\hat{\bm{a}} = R(\Phi,\bm{f},\{k_1,\dots,k_i\}/\{k_i\}),\\
\bm{\bar{z}}_j = \Phi R(\Phi,\bm{f},\{k_1,\dots,k_i\}/\{k_i\}),
\end{cases}
\end{equation}
and updating $S\rightarrow S/\{j\}$ (that is, remove $j$ from the set $S$) if $I(\Phi R(\Phi,\bm{f},S);\bm{f}|\bm{\bar{z}}_j)\approx 0$. The result of the backward ER is the reduced set of indices $S=\{\ell_1,\dots,\ell_n\}$ with $n \leq m$, together with the corresponding parameters $a_{\ell_1},\dots,a_{\ell_n}$ ($a_j=0$ for $j\not\in S$) computed as $\bm{a}=R(\Phi,\bm{f},S)$, and accordingly the recovered model $\bm{f}\approx \bm\phi\bm{a}=\bm\phi_S\bm{a}_S=a_{\ell_1}\bm\phi_{\ell_1}+\dots+a_{\ell_n}\bm\phi_{\ell_n}$. }
In practice, mutual information and conditional mutual information need to be estimated from data, and whether or not the estimated values should be regarded as zero is typically done via (approximate) significance testing, the details of which are provided in {\it Methods} (also see Sec.~\ref{Sec:SI_EntropicRegression}).

\begin{algorithm}
\caption{Entropic Regression}\label{alg:ER}
\begin{algorithmic}[1]
\Procedure{Initialization:}{$\bf{f},\Phi$}
\State Tolerance ($tol$) Estimation.
\State For a set of index $S$, define the function $R(\Phi,\bm{f},S)=\Phi_S^\dagger\bm{f}$
\EndProcedure

\Procedure{Forward ER:}{$\bf{f},\Phi,tol$}
\State $S_f = \emptyset , p = \emptyset, v = \infty, z = \emptyset$
\While{$v > tol$}
\State  $S_f \leftarrow p$
\State  $I^{est}_{j} := I(\Phi R(\Phi,\bm{f},\{S_f,j\});\bm{f} | z ) $. for all $j \notin S_f$
\State  $v$, $p$ $:= \max_{j} I^{est}_{j}$
\State $\hat{\bm{a}} := R(\Phi,\bm{f},\{S_f, p\}))$
\State $z := \Phi \hat{\bm{a}}$
\EndWhile
\State \textbf{return} $S_f$
\EndProcedure

\Procedure{Backward ER:}{$\bf{f},\Phi,tol,S_f$}
\State $S_b = S_f , p = \emptyset, v = - \infty$
\While{$v < tol$}
\State  $S_b := \{S_b \} - \{p\}$
\ForAll{$j \in S_b $}
\State $\hat{\bm{a}} := R(\Phi,\bm{f},\{S_b\} - \{j\}))$
\State $z := \Phi \hat{\bm{a}}$
\State $I^{est}_{j} := I(\Phi R(\Phi,\bm{f},{S_b}) ; \bm{f} | z )$,
\EndFor
\State  $v$, $p$ $:= \min_{j} (I^{est}_{j})$
\EndWhile\label{BWwhile}
\State \textbf{return} $S_b$
\EndProcedure

\State \textbf{return} $S = S_b.$ 
\end{algorithmic}
\end{algorithm}

\subsubsection*{Numerical Experiments: Outliers, Expansion Order, and the Paradox of Sparsity}
To demonstrate the utility of ER for nonlinear system identification under noisy observations, we benchmark its performance against existing methods including least squares (LS), orthogonal least squares (OLS), Lasso,
\textcolor{black}{as well as SINDy and its extension by Tran and Ward (TW).}
The details of the existing approaches are described in the Methods Section. 
The examples we consider represent different types of systems and scenarios, including both ODEs and PDEs.
In addition, we consider different noise models and especially the presence of outliers in order to evaluate the robustness of the respective methods.

For each example system, we sample the state of each variable at a uniform rate of $\Delta{t}$ to obtain a multivariate time series $\{\bm{z}(t_i)\}_{k=1,\dots,N;i=1,\dots,\ell}$ where $\bm{z}=[z_1,\dots,z_d]^\top\in\mathbb{R}^d$; then we add noise to each state variable  and obtain the noisy empirical time series denoted by $\{\hat{\bm{z}}(t_i)\}$, where
\begin{equation}
\hat{z}_k(t_i) = z_k(t_i) + \eta_{ki},
\end{equation}
with $\eta_{ki}$ \textcolor{black}{representing state observational} noise. \textcolor{black}{The vector field $\bm{F}$ is estimated using central difference on the noisy time series $\{\hat{\bm{z}}(t)\}$.}

\medskip\noindent
\textbf{Example 1. Chaotic Lorenz system.}
Our first detailed example data set was generated by noisy observations from a chaotic Lorenz system, which is represented by a three-dimensional ODE  which is a prototype system as a minimal model for thermal convection obtained by a low-ordered modal truncation of the Saltzman PDE~\cite{Saltzman1962}, and for many parameter combinations exhibits chaotic behavior~\cite{Lorenz1963}. In our standard notation, we have $\bm{z}=[z_1,z_2,z_3]^\top$ and
\begin{equation}\label{eq:Lorenz}
\begin{cases}
\dot{z}_1=F_1(\bm{z})=\sigma(z_2-z_1), \nonumber \\ 
\dot{z}_2=F_2(\bm{z})=z_1(\rho-z_3)-z_2, \nonumber \\ 
\dot{z}_3=F_3(\bm{z})=z_1z_2-\beta z_3,
\end{cases}
\end{equation}
with default parameter values $\sigma=10$, $\rho=28$ and $\beta=8/3$ unless otherwise specified. We consider a standard polynomial basis as in Eq.~\eqref{eq:polybasis}.
\textcolor{black}{Over recent years, the Lorenz system has become a favorable and standard example for testing SID methods and typically requires tens of thousands of measurements for accurate reconstruction~\cite{TranWard, Brunton2015}.}

\begin{figure}
    \centering
    \includegraphics[width=0.8\textwidth]{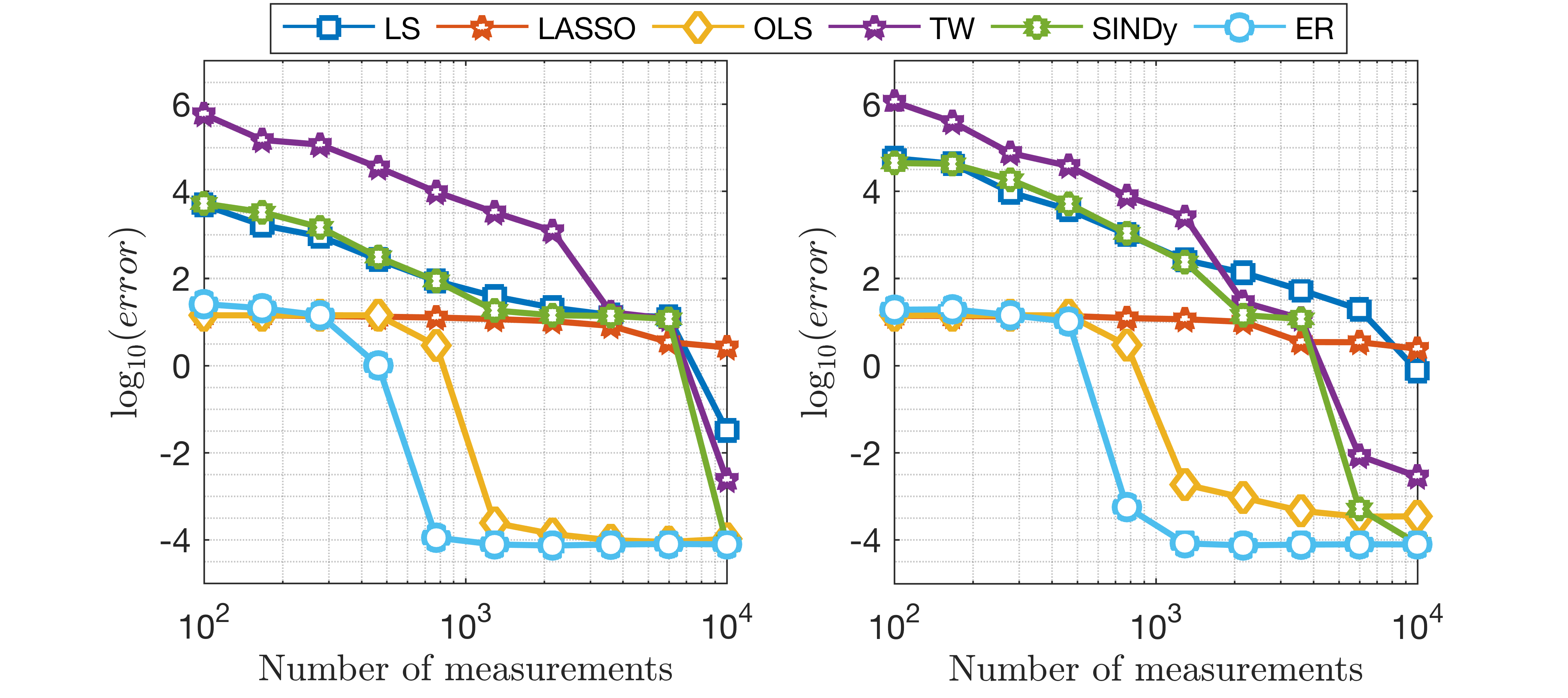}
    \caption{\small Lorenz system. We perform 100 runs for the comparison, no outliers, 0.0005 step size, and we considered the median result out of 100 runs.  The figure shows the error in the parameter estimation for a Lorenz system but subject to noisy measurements by Gaussian noise, with $\epsilon = 10^{-4}$, and using a $5^{th}$-order polynomial expansion. We see that ER and OLS has an overall superior performance compared to others standard methods. We see that SINDy, and TW are less successful (\textcolor{black}{under large span of tuning parameters, see Fig.(\ref{fig:TWcontourplot})}) at this number of measurements even with low noise levels.}
    \label{fig:ER_Lorenz}
\end{figure}

First, we compare several nonlinear SID methods in reconstructing the Lorenz system \textcolor{black}{when the state observational noise is drawn independently from a Gaussian distribution, $\eta\sim\mathcal{N}(0,\epsilon^2)$. As we discussed before, this translates into effective noise that is not necessarily Gaussian or even independent.}
\textcolor{black}{Fig.~\ref{fig:ER_Lorenz} shows the error in the estimated parameters where, $error = \Vert \bm{a}_{true} - \bm{a}_{estimated} \Vert_2$.}
\noindent \textcolor{black}{As shown in  Fig.~\ref{fig:ER_Lorenz}, even with observational noise as low as $\epsilon=10^{-4}$, ER and OLS outperform all other methods. In this low noise regime, SINDy required more measurements (around 4 times) to reach similar accuracy as ER. 
In comparison, as noted in \cite{TranWard, Brunton2015} and in the implementation provided by the authors, for SINDy and TW methods to yield accurate reconstruction the number of measurements is at the order of $10^4$.} 

\begin{figure}
    \centering
    \includegraphics[scale=0.7]{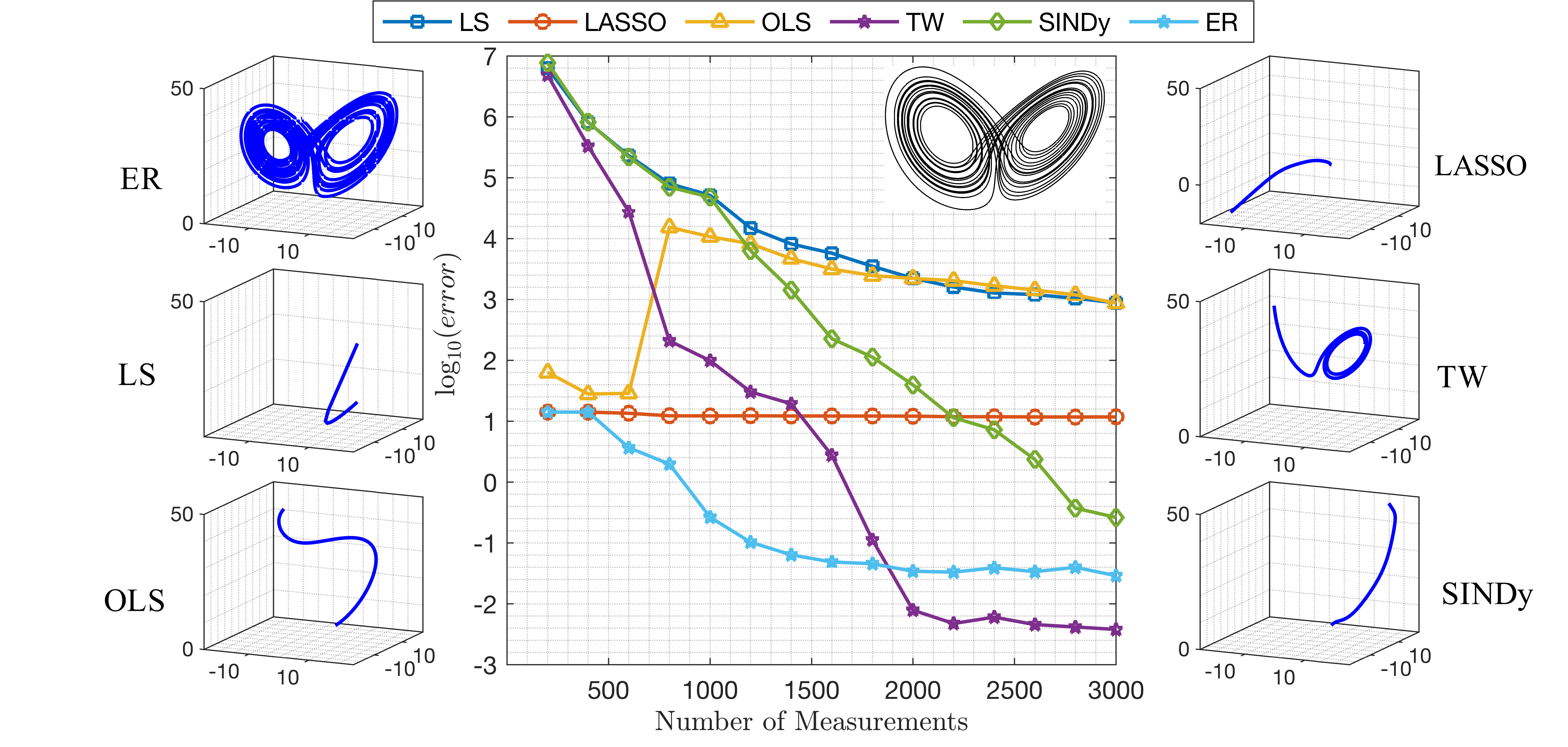}
    \caption{SID for the Lorenz system when the observations are corrupted by outliers.  Contrast to Fig.~\ref{fig:ER_Lorenz}. As before, we specify a level of persistent Gaussian observation noise, $\eta\sim\mathcal{N}(0,\epsilon_1)(1-Ber(p))$, but now furthermore we allow for an ``outlier noise'', as ``occasional" bursts of much larger perturbations, $\eta\sim\mathcal{N}(0,\epsilon_1 + \epsilon_2)Ber(p)$, where $Ber(p)$ is the standard Bernoulli random variable (0 or 1 with probability ratio $p$, and $0\leq p \leq 1$). \textbf{(Middle)} Error in estimated parameters for Lorenz system given in Eq.~\ref{eq:Lorenz} with noise, $\epsilon_1 = 10^{-5}$, $\epsilon_2 = 0.2$, $5^{th}$-Order polynomial expansion, and $p=0.2$.  Lorenz system dynamics is shown in the upper right corner. We see that ER has fast convergence at a low number of measurements, followed by TW which required twice number of measurements. Different from TW, in our ER method we focus in detecting the true sparse structure with the presence of outliers, without any attempts to neglect outliers based on some weight function to achieve higher accuracy which is the case in TW method. This point clearly appears in Fig.(\ref{fig:exactRecovery} where we see that although TW achived higher accuracy, it has low exact recovery probability, while ER reached exact recovery probability more than 90\%.  \textcolor{black}{A detailed statistics box-plot (quartiles, median,...,etc) over the 100 runs with 1500 measurements is shown in Fig.~(\ref{fig:LorenzBoxPlot}).} \textbf{(Side panels)} \textcolor{black}{Typical trajectories generated by the reconstructed dynamical systems, where for each method we show results using the ``median" solution, that is, recovered system whose corresponding parameter estimation error is at the median over a large number of independent simulation runs. In each such simulation, 1500 samples are used.
    Comparing with the true Lorenz attractor (upper right corner in the main panel), we see that the only reasonable reconstruction in this case was produced by ER.}}
    \label{fig:ER_Lorenz2}
\end{figure}

Next, to explore the performance of SID methods under the presence of outliers, we conduct additional numerical experiments.
The extent to which outliers present is controlled by a single parameter $p$: each observation is subject to an added noise $\eta$, where $\eta\sim\mathcal{N}(0,\epsilon_1^2)$ with probability $1-p$ and $\eta\sim\mathcal{N}(0,\epsilon_1^2+\epsilon_2^2)$ with probability $p$. \textcolor{black}{Here we use $\epsilon_1 = 10^{-5}$, $\epsilon_2 = 0.2$ and $p=0.2$}. The results of SID are shown in Fig.~\ref{fig:ER_Lorenz2}. 
\textcolor{black}{Compared to Fig.(\ref{fig:ER_Lorenz}), we see that with $p > 0$ OLS performance drops due to the increasing occurrence of large noise and outliers whereas ER remains its capacity of accurately identifying the underlying system.} 
\textcolor{black}{As an example, in each of the side panels of Fig.~\ref{fig:ER_Lorenz2}, we show the trajectory of the identified dynamics using the median solution of each method. It is clear that under such noisy chaotic dynamics and at a relatively under-sampled regime, ER method successfully recovers the system dynamic. As an ample amount of data becomes available, we note that TW method starts to produce excellent reconstruction which is consistent with recent findings reported in Ref.~\cite{TranWard}.}

\begin{figure}
    \centering
    \includegraphics[scale=1]{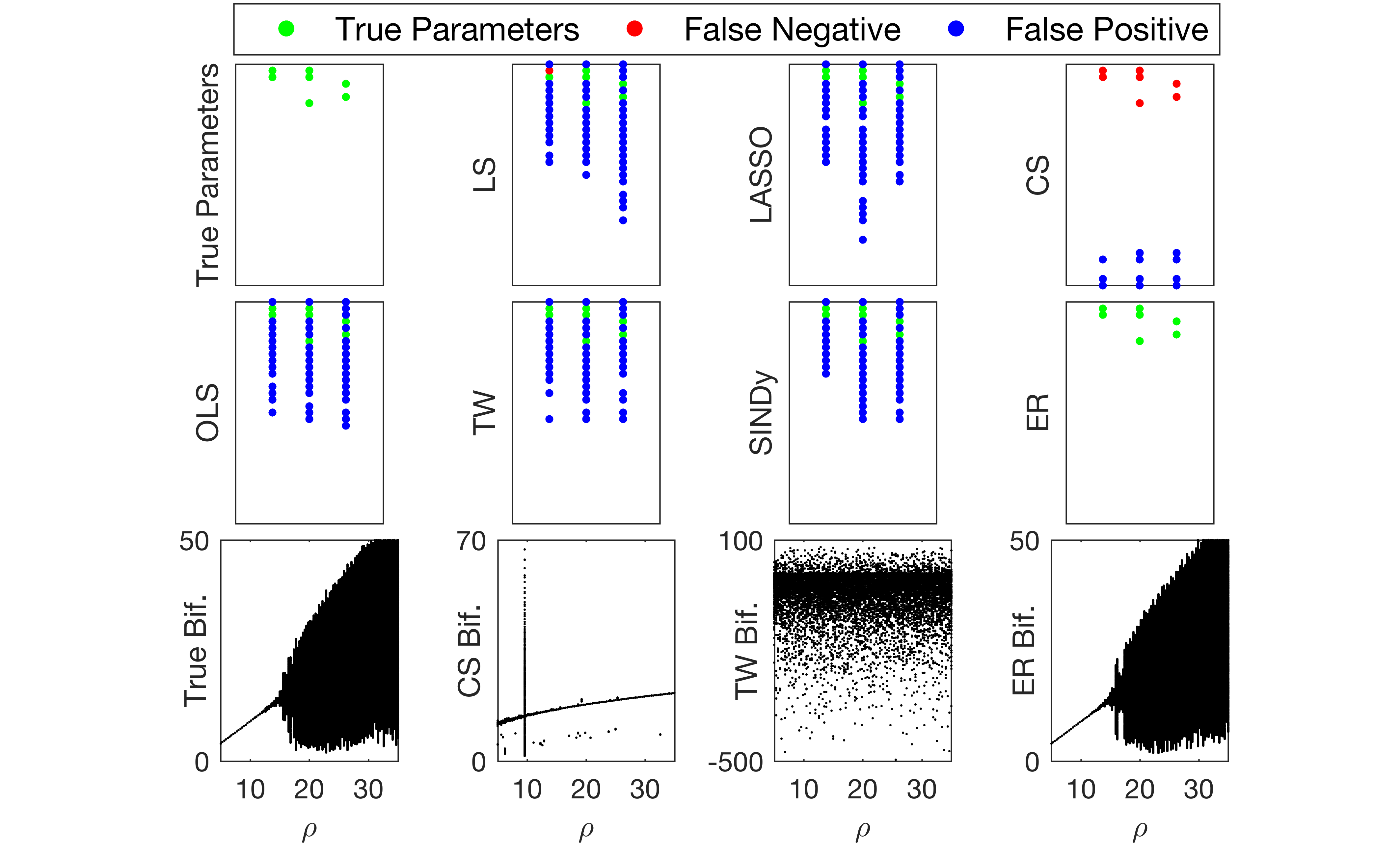}
    \caption{Sparse representation of the solution found by solvers using 1500 measurements, and $p = 0.2$ on Fig.(\ref{fig:ER_Lorenz2}). The upper left corner shows the true solution of Lorenz system. The bottom column shows the bifurcation diagram on $z$ dimension of Lorenz system with $\rho \in [5, 30]$ as bifurcation parameter, created using 5000 initial conditions evolved according the recovered solution.}
    \label{fig:ER_Lorenz3}
\end{figure}

Given that a major theme of modern SID is to seek for {\it sparse} representations, and the Lorenz system under standard polynomial basis is indeed sparse, it is worth asking: what are the respective structure identified by the different methods? In Fig.~\ref{fig:ER_Lorenz3} we compare the structure of the identified model using different methods across a range of parameter values for $\rho$. \textcolor{black}{In this case, under the presence of large noise and outliers ($p=0.2$), none of the methods examined here, including recently proposed sparsity-promoting (CS, SINDy) and outlier-resilient (TW) methods, is able to identify the correct structure. The proposed ER method, however, does identify the correct structure. It is worth pointing out that, often times when expressed in the right basis, a model will appears to be sparse, the converse is not true: just because a method return a sparse solution does not suggest (at all) the such a solution gives a reasonable approximation of the true model structure. Interestingly, as we show in Fig.~\ref{fig:CS357} and Fig.~\ref{fig:CS7}, for the same system and data, as more basis functions are used--that is, when the true dynamics becomes sparser--the reconstructed dynamics using existing methods (such as CS) can become worse.}



\medskip\noindent
\textbf{Example 2. Kuramoto-Sivashinsky equations.}
To further demonstrate the power of ER, we consider a nonlinear PDE, namely the Kuramoto-Sivashinsky (KS) equation~\cite{Kuramoto1976,Kuramoto1978,Sivashinsky1977,Hyman1986,Lan2008}, which arises as a description of flame front flutter of gas burning in a cylindrically symmetric burner. It has become a popular  example of a PDE that exhibits chaotic behavior, in particular spatiotemporal chaos~\cite{Christiansen1997, Hohenberg1989}. We will consider Kuramoto-Sivashinsky system in the following form,
\begin{equation}
\label{eq:KSE}
u_t = -\nu u_{xxxx} - u_{xx} + 2uu_x,\quad\quad (t,x)\in
[0,\infty)\times (0,L)
\end{equation}
in periodic domain, $u(t,x) = u(t,x+L)$, and we restrict our solution to the subspace of odd solutions $u(t,-x) = -u(t,x)$.
The viscosity parameter $\nu$ controls the suppression of solutions with fast spatial variations, and is set to $\nu = 0.029910$ under which the system exhibit chaotic behavior \cite{Christiansen1997}. 

Since a PDE corresponds to an infinite-dimensional dynamical system, in practice we focus on an approximate finite-dimensional representation of the system, for example, by Galerkin-projection onto basis functions as infinitely many ODE's in the corresponding Banach space.

To develop the Galerkin projection, we follow the procedure as presented in~\cite{Artuso2002},
to expand a periodic solution $u(x,t)$ using a discrete spatial Fourier series,
\begin{equation}\label{eq:KSE_Fourier}
u(x,t) = \sum_{-\infty}^{\infty} b_{k}(t)e^{ikqx}, \hspace{1cm} \text{where} \hspace{0.2cm} q=\frac{2\pi}{L}.
\end{equation}
Notice that we have written this Fourier series of basis elements $e^{ikqx}$ in terms of time varying combinations of basis elements. For simplicity, consider $L=2\pi$, then $q=1$ for the following analysis.  This is typical \cite{robinson2001infinite} with the representation of a PDE as infinitely many ODE's in the Banach space, where orbits of these coefficients therefore become time varying patterns by Eq.~(\ref{eq:KSE_Fourier}).
Substituting Eq.~(\ref{eq:KSE_Fourier}) into Eq.~(\ref{eq:KSE}), we produce the infinitely many evolution equations for the Fourier coefficients,
\begin{equation}
\dot{b}_{k} = (k^2 - \nu k^4)b_k + ik \sum_{m=-\infty}^{\infty} b_{m}b_{k-m}
\end{equation}

In general, the coefficients $b_k$ are complex functions of time $t$.  However, by symmetry, we can reduce to a subspace by considering the special symmetry case that $b_k$ is pure imaginary, $b_k = ia_k$ and  $a_k\in {\mathbb R}$.  Then,
\begin{equation}\label{eq:KSE_ak}
\dot{a}_{k} = (k^2 - \nu k^4)a_k - k \sum_{m=-\infty}^{\infty} a_{m}a_{k-m}.
\end{equation} 
 where $k =1,..,N_m$. 
 However, the assumption that there is a slow manifold (slow modes as an inertial manifold \cite{robinson2001infinite,Jolly1990,Ramdani2000,Jolly2001}) suggests the practical matter that a finite truncation of the series Eq.~(\ref{eq:KSE_Fourier}), and correspondingly the a reduction to finitely  many  ODEs will suffice. Therefore we choose a sufficiently large number of modes $N_m$.  Then we solve the resulting  $N_m$-dimensional ODE (\ref{eq:KSE_ak}) to produce the estimated solution of $u(x,t)$ by (\ref{eq:KSE_Fourier}), and use such data for the purpose of SID, have meaning to estimate the structure and parameters of the ODE model (\ref{eq:KSE_ak}).

\begin{figure}[hbtp]
\centering
\includegraphics[scale=1]{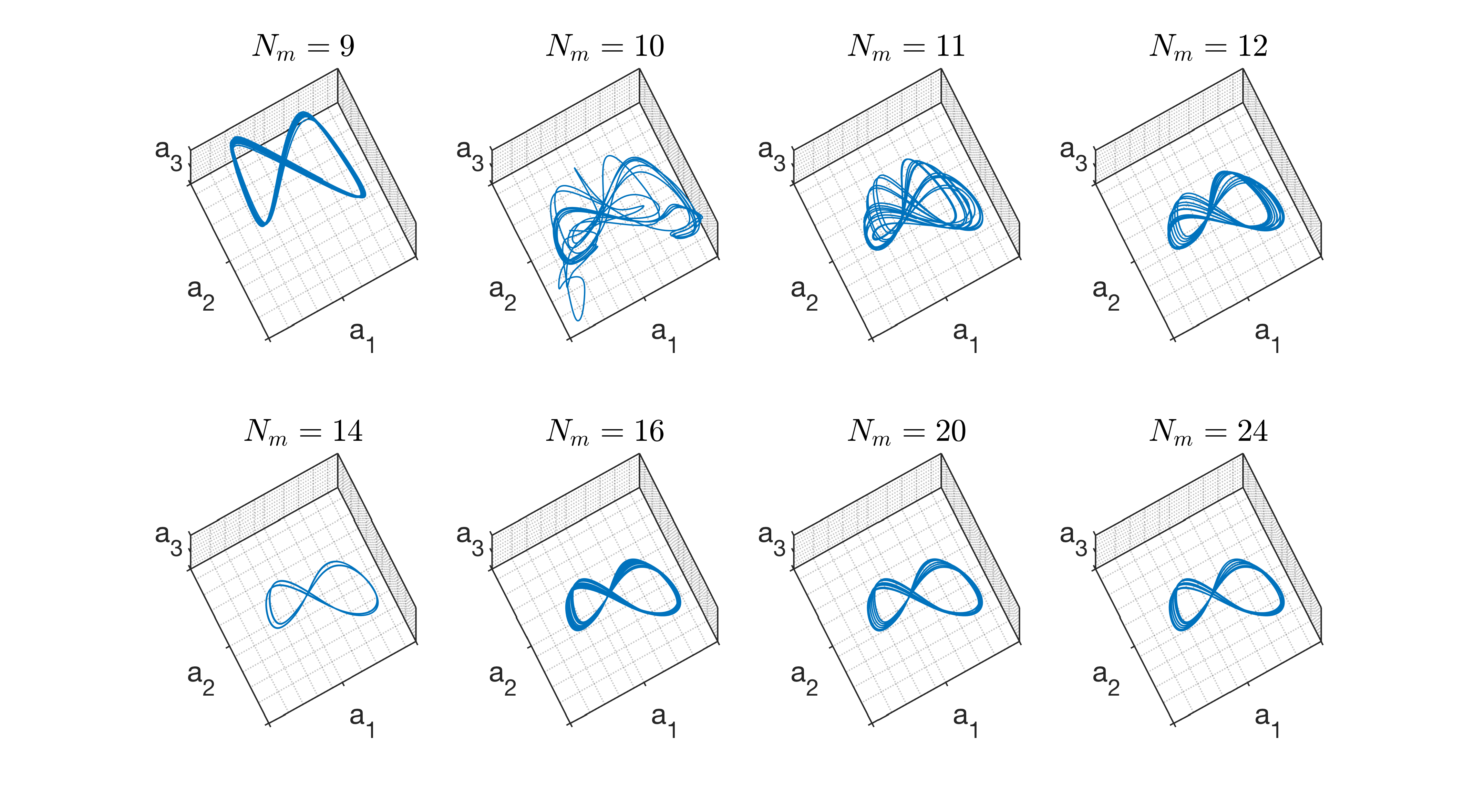}
\caption{The first three modes of the ODE Eq.(\ref{eq:KSE_ak}) solution. We show the modes $a_1, a_2$ and $a_3$ for selected number of modes. For clear view, we fixed the axis limits to be $a_1 \in [-1.21, 1.06], a_2 \in [-0.75, 0.98]$ and $a_3 \in [-1.1, 1.12]$ for all plots. We found that there was no significant addition to the dynamic with $16 < N_m$. (meaning that $N_m = 16$ was enough to describe the system).}
\label{fig:KSE_modes}
\end{figure}


\textcolor{black}{Fig.~\ref{fig:KSE_modes}  shows the first three dimensions plot under different number of modes. We see that using just a few number of modes ($N_m = 8,...,11$) is insufficient to capture the true dynamical behavior of the system whereas too large a number of modes ($N_m = 20,24$) may be unnecessary. In this example, an adequate but not excessive number of modes seems to be around $N_m=16$, as no significant information is gained by increasing $N_m$.}

\begin{figure}[hbt]
\centering
\includegraphics[scale=0.9]{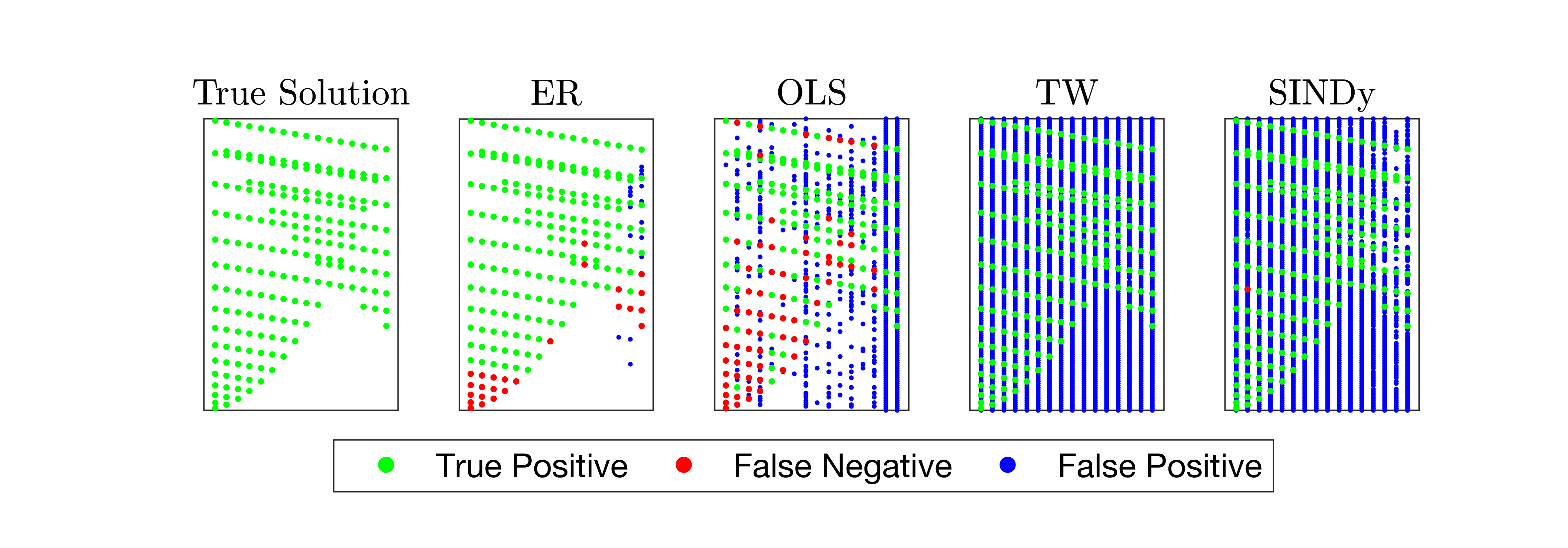}
\caption{In analogy to Fig.~\ref{fig:ER_Lorenz3}, sparse representation of KSE solution by different methods. CS, LASSO have been excluded for their high computation complexity.}
\label{fig:KSE_sparse}
\end{figure}

\textcolor{black}{Fig.~\ref{fig:KSE_sparse} shows the sparse structure of the recovered solution by different methods. Here we mention that the true non-zero parameters of KSE using $N_m = 16$ are 200 parameters that vary in the magnitude from 0.9701 to 1705. With the second order expansion, our basis matrix will have 153 candidate functions, and it will be nearly singular with condition number $4\times 10^{7}$. Likely due to such high condition number, neither TW nor SINDy gives reasonable reconstruction. In particular, we note that the solution of SINDy is already optimized by selecting the threshold value $\lambda$ that is slightly above $\lambda_*$ where here $\lambda_*\approx0.1731$ is the smallest magnitude of the true nonzero parameter of the full least squares solution. A larger value of $\lambda$ only worsens the reconstruction, as we found numerically.}

\begin{figure}
    \centering
    \includegraphics[scale=2]{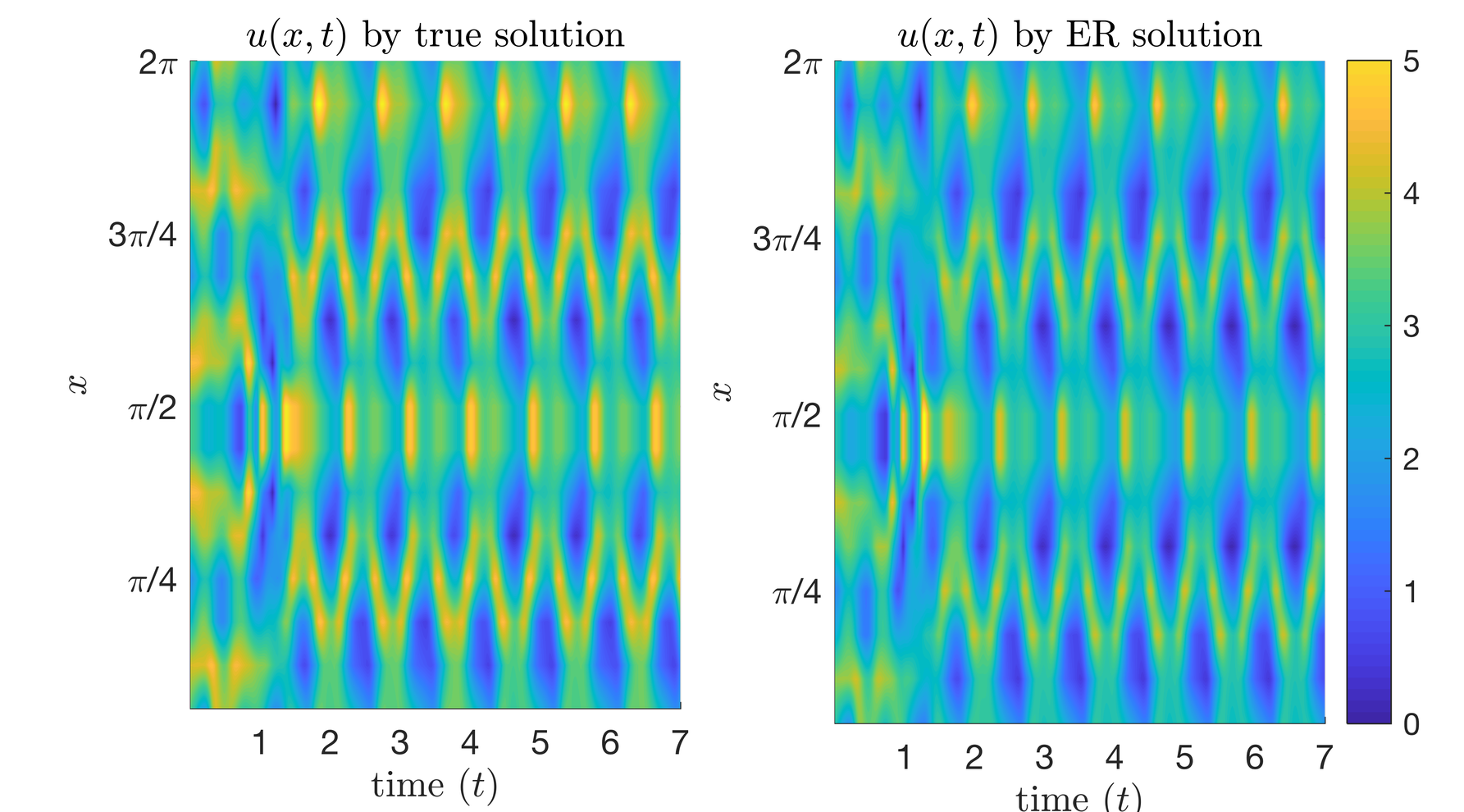}
    \caption{$u(x,t)$ constructed by the true solution (left) and the ER solution (right) using Eq.(~\ref{eq:KSE_Fourier}). OLS and TW was not able to re-produce the dynamic and they diverge after few iterations. we see that the reconstructed dynamic using ER solution is identical to the true solution with a minor difference in the transient time, although there was a false negative in the ER solution. ER detected the stiff parameters that dominate the overall dynamic. Sloppiness of some KSE parameters make there influence practically negligible to the overall dynamic.}
    \label{fig:uxt}
\end{figure}

The OLS method overcomes the disadvantage of LS by iteratively finding the most relevant ``feature'' variables, where relevance is measured in terms of (squared) model error; but it comes at a price: similar to LS, the OLS is sensitive to outliers in the data and such sensitivity seems to be even more amplified due to the smaller number of terms typically included in OLS as compared to LS, \textcolor{black}{which cause the high false negative rate in the OLS solution}. 
\textcolor{black}{Although ER solution has few false negatives, but was completely able to recover the overall dynamic of the system as shown in Fig.(~\ref{fig:uxt}), while all other solutions diverges and failed to recover $u(x,t)$.}

\medskip\noindent
\textbf{Example 3. Double Well Potential.} Finally, in order to gain further insights into why standard methods fail under the presence of outliers, we consider a relatively simple double-well system, with 
\begin{equation}\label{eq:doubleWell}
f(x) = x^4 - x^2.
\end{equation}
Suppose that we measure $x$ and $f$, can we identify the function $f(x)$?
We sample 61 equally spaced measurements for $x \in [-1.2, 1.2]$, and we construct $\Phi$ using the $10^{th}$ order polynomial expansion with $K = 11$ is the number of candidate functions. Then, we consider a single fixed value corrupted measurement to be $f(0.6) = 0.2$.

\begin{figure}[ht]
\centering
\includegraphics[scale=0.3]{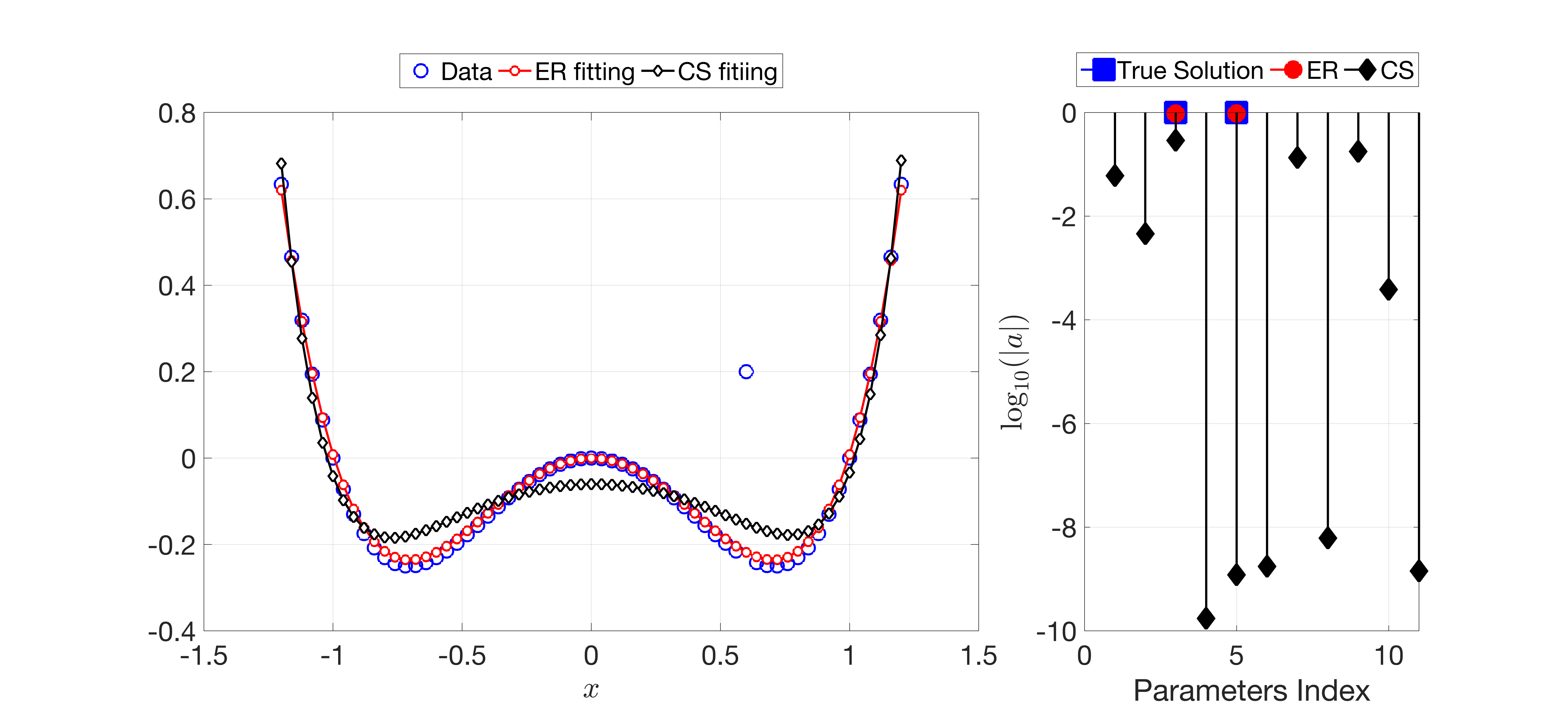}
\caption{Double well potential given by Eq.~(\ref{eq:doubleWell}) data fitting using ER and CS. CS solution found as the solution with minimum residual from 100 log-spaced values of $\epsilon \in [10^{-9}, 10^{2}]$. }
\label{fig:dw}
\end{figure}

Fig.~\ref{fig:dw} shows the results the double-well SID under a single outlier in the observation. We see the robustness of ER solution to the outliers while CS failed in detecting the system sparse structure. For the sake of clearness,  Fig.~\ref{fig:dw} shows the results for CS and ER. The results for each solver and details are provided in Sec.~(\ref{sec:SI_doublewell}) in addition to more numerical examples.

\section*{Discussion}
The main theme of the paper is on nonlinear system identification (SID) \textcolor{black}{under noisy observations, which is to learn the functional form and parameters of a nonlinear system based on observations of its states under the presence of noise and outliers.}
We recast the problem into the form of an inverse problem using a basis expansion of the nonlinear functions. Such basis expansion, however, renders the resulting problem inherently high dimensional even for low-dimensional systems. In practice, the need for finite-order truncation as well as the presence of noise causes additional challenges.
\textcolor{black}{For instance, even under iid Gaussian observational noise for the state variables, the effective noise in the inverse problem is not necessarily so.} As we demonstrate using several example systems, including the chaotic Lorenz system and the Kuramoto-Sivashinsky equations, existing SID methods are prone to noise, and can be quite sensitive to the presence of outliers. We identify the root cause of such non-robustness being the metric nature of the existing methods, as they quantify error based on metric distance, and thus a handful of data points that are ``corrupted" by large noise can dominate the model fit. Each of the existing methods we considered has this property, which includes the least squares, compressive sensing, and Lasso. From a mathematical point of view, each method can be interpreted as a functional that maps input data to a model, through some optimization process. In a noisy setting, the output model should ideally change smoothly with respect to the input data, not just continuously. Our results suggest that these popular methods in fact do suffer from a sensitive dependence on outliers, as a few corrupted data can already produce very poor model estimates. Alarmingly, the now-popular CS method, which is based on sparse regression, can force to select a completely wrong sparse model under noisy input data, and this occurs even when there is just a single outlier. This is by no means contradicting previous findings of the success of CS in SID, as in such work noise is typically very small, and here we are considering a perhaps more realistic scenario with larger noise.

To fill the vacancy of SID methods that can overcome outliers, we develop an information-theoretic regression technique, called entropic regression (ER), that combines entropy measures with an iterative optimization for nonlinear SID. We show that ER is robust to noise and outliers, in the otherwise very challenging circumstances of finding a model that explains data from dynamical stochastic processes. The key to ER's success is its ability to recover the correct and true sparsity structure of a nonlinear system under basis expansions, despite either relatively large noise, or alternatively even relatively many even larger outliers. In this sense ER is superior to any other method that we know of for such settings. \textcolor{black}{Note that in the ER algorithm, least squares is used to estimate the parameters of those basis functions that are deemed relevant where relevance is detected using an information-theoretic measure that is insensitive to noise and outliers. The choice of least squares in the regression step in ER is not necessarily an optimal choice and can be potentially replaced by more advanced methods (e.g., those developed in robust regression). In the current implementation of ER we adopted least squares mainly due to its computational advantage over alternative methods.}
On a more fundamental level, ER's robustness against outliers may likely be attributed to an important principle in information theory called the asymptotic equipartition property (AEP) \cite{Cover2005}.  The outcome of this principle is that sampled data can be partitioned into ``typical'' samples and ``atypical'' samples, with the rare atypical samples end up influencing the estimated entropy relatively weakly. Since ER measures relevance by entropy  instead of metric distance, a few outliers, no matter how far away they are from the rest of the data points, tend to have minimal impact on the model identification process.
So the general interpretion we make here is that outliers observations are likely atypical, but not part of the core of data that carry the major estimation of the entropy.  This foundational concept of information theory is likely the major source of robustness of our ER method to system identification.

\section*{Methods}



\subsection*{Existing metric-based methods for system identification}
Recall (from the main text) that we recast the nonlinear system identification problem here. Given a truncated basis representation of each component of the vector field $\bm{F}$, expressed as
\begin{equation}
F_i(\bm{z}) = \sum_{k=0}^{K}a_{ik}\phi_k(\bm{z}),
\end{equation}
we consider sampled data $\hat{\bm{z}}$ and the estimated vector field $\hat{\bm{F}}$, from which the coefficients (parameters) $\{a_{ik}\}$ are to be determined. In general, we use subscript ``$t$'' to index the sampled data, and thus the $t$-th sample satisfies the equation
\begin{equation}
\hat{F}_i(\hat{\bm{z}}(t)) = \sum_{k=0}^{K}a_{ik}\phi_k(\hat{\bm{z}}(t)) + \xi_i(t),~(t=1,\dots,T;i=1,\dots,n).
\end{equation}
Here $\xi_i(t)$ \textcolor{black}{is the effective noise that represents the accumulative impact of truncation error, state observational noise as well as approximation error in the estimation of derivatives. Consequently, an iid Gaussian noise additive to the states $\bm{z}_i(t)$ can translate into correlated non-Gaussian effective noise for $\xi_i(t)$.}

Having transformed a system identification problem into an parameter estimation problem (or inverse problem) in the form of 
\begin{equation}\label{eq:inverse_i_rep}
\bm{f}^{(i)} = \Phi\bm{a}^{(i)} + \bm\xi^{(i)},
\end{equation}
where $\bm{f}^{(i)}=[\hat{F}_i(\hat{\bm{z}}(1)),\dots,\hat{F}_i(\hat{\bm{z}}(T))]^\top\in\mathbb{R}^{T\times1}$ represents the estimated function $F_i$ ($i$-th component of the vector field $\bm{F}$), $\Phi=[\bm{\phi}^{(1)},\dots,\bm{\phi}^{(K)}]\in\mathbb{R}^{T\times K}$ (with $\bm{\phi}^{(k)}=[\phi_k(\hat{\bm{z}}(1)),\dots,\phi_k(\hat{\bm{z}}(T))]\in\mathbb{R}^{T\times1}$) represent sampled data for the basis functions,
$\bm\xi^{(i)}=[\xi_i(1),\dots,\xi_i(T)]^\top\in\mathbb{R}^{T\times1}$ represents \textcolor{black}{effective} noise, and
$\bm{a}^{(i)}=[a_{i1},\dots,a_{iK}]^\top\in\mathbb{R}^{K\times1}$ is the vector of parameters which is to be determined.
Since the form of the equation~\eqref{eq:inverse_i_rep} is the same for each $i$, we omit the index when discussing the general methodology, and consider the following linear inverse problem
\begin{equation}\label{eq:inverse_rep}
\bm{f} = \Phi\bm{a} + \bm\xi,
\end{equation}
where $\bm{f}\in\mathbb{R}^{T\times1}$ and $\Phi\in\mathbb{R}^{T\times K}$ are given, with the goal is to estimate $\bm{a}\in\mathbb{R}^{K\times1}$ \textcolor{black}{when the effective noise is not necessarily from independent multivariate Gaussian distribution.}

\subsubsection*{Least Squares (LS)}
The most commonly used approach to estimate $\bm{a}$ in Eq.~\eqref{eq:inverse_rep} is to use the least squares criterion, which finds $\bm{a}$ by solving the following least squares minimization problem:
\begin{equation}
	\min_{\bm{a}\in\mathbb{R}^K}\|\Phi\bm{a}-\bm{f}\|_2.
\end{equation}
The solution can be explicitly computed, giving
\begin{equation}
	\bm{a}_{\mbox{\tiny(LS)}}=\Phi^\dagger\bm{f},
\end{equation}
where $\Phi^\dagger$ denotes the pseudoinverse of the matrix $\Phi$~\cite{Golub}. Note that in the special case where the minimum is zero (which is unlikely under the presence of noise), the minimizer is not unique and the ``least-squares" solution typically refers to a vector $\bm{a}$ that has the minimal $2$-norm and solves the equation $\Phi\bm{a}=\bm{f}$.
The LS method has several advantages: it is analytically traceable and easy to solve computationally using standard linear algebra routines (e.g., SVD).
However, a main disadvantage of the LS approach in system identification, as we discuss in the main text, is that it generally produces a \textcolor{black}{``dense" solution, where most (if not all) components of $\bm{a}$ are nonzero, which is a severe overfitting of the actual model. 
This (undesired) feature also makes the method sensitive to noise, especially in the under-sampling regime.}

\subsubsection*{Orthogonal Least Squares (OLS)}
In orthogonal least squares (OLS)~\cite{Chen1989,Wang1995,Korenberg1988}, the idea is to iteratively select the columns of $\Phi$ that minimize the \textcolor{black}{(2-norm)} model error, which corresponds to iterative assigning nonzero values to the components of $\bm{a}$. In particular, the first step is to select basis $\bm\phi_{k_1}$ and compute the corresponding parameter $a_{k_1}$ and residual $\bm{r}_{1}$ according to
\begin{equation}
\begin{cases}
(k_1,a_{k_1})=\argmin_{k,c}\|\bm{f}-c\bm\phi_k\|_2,\\
\bm{r}_{1}=\bm{f}-\bm\phi_{k_1}a_{k_1}.
\end{cases}
\end{equation}
Then, one iteratively selects additional basis functions (until stopping cretia is met) and compute the corresponding parameter value and residual, as
\begin{equation}
\begin{cases}
(k_{\ell+1},a_{k_{\ell+1}})=
\argmin_{k,c}\|\bm{r}_\ell-c\bm\phi_k\|_2,\\
\bm{r}_{\ell+1}=\bm{r}_\ell-\bm\phi_{k_{\ell+1}}a_{k_{\ell+1}}.
\end{cases}
\end{equation}
\textcolor{black}{As for stopping criteria, there are several choices including AIC and BIC. In this work, in the absence of knowledge of the error distribution, we adopt a commonly used criterion where the iterations terminate when the norm of the residual is below a prescribed threshold.  To determine the threshold, we consider 50 log-spaced candidate values in the interval $[10^{-6},100]$ and select the best using 5-fold cross validation.}

\subsubsection*{Lasso}
A principled way to impose sparsity on the model structure is to explicitly penalize solution vectors that are non-sparse, by formulating a regularized optimization problem:
\begin{equation} \label{eq:LASSO}
\min_{\bm{a}\in\mathbb{R}^K}
\left(\|\Phi\bm{a}-\bm{f}\|_2^2+\lambda\|\bm{a}\|_1\right),
\end{equation}
where the parameter $\lambda\geq0$ controls the extent to which sparsity is desired: as $\lambda\rightarrow\infty$ the second term dominates and the only solution is a vector of all zeros, whereas at the other extreme $\lambda=0$ and the problem becomes identical to a least squares problem which generally yields a full (non-sparse) solution. Values of $\lambda$ in between then balances the ``model fit'' quantified by the 2-norm and the sparsity of the solution characterized by the 1-norm. For a given problem, the parameter $\lambda$ needs to be tuned in order to specify a particular solution. A common way to select $\lambda$ is via cross validation~\cite{Hastie2015}. In our numerical experiments, we choose $\lambda$ span according to ~\cite{Hastie2015}, with 5-Folds cross validation and 10 values $\lambda$ span. We adopt the CVX solver~\cite{Grant2008}, and from all the solutions found for each $\lambda$ we select the solution with minimum residual.

\subsubsection*{Compressed sensing (CS)}
Originally developed in the signal processing literature~\cite{Candes2006a,Candes2006b,Donoho2006}, the idea of compressed sensing (CS) has been adopted in several recent work in nonlinear system identification~\cite{Wang2011,Wang2016}
Under the CS framework, one solves the following constrained optimization problem,
\begin{equation} \label{eq:CS}
\begin{cases}
\argmin_{\bm{a}}\|\bm{a}\|_1,\\
\mbox{subject to}~\|\Phi\bm{a}-\bm{f}\|\leq\epsilon,
\end{cases}
\end{equation}
where the parameter $\epsilon\geq0$ is used to relax the otherwise strict constraint $\Phi\bm{a}=\bm{f}$, to allow for the presence of noise in data. In our numerical experiments, we choose 10 log-spaced values for $\epsilon \in [10^{-6}, 100]$, and 5-Folds cross validation. We adopt the CVX solver~\cite{Grant2008}, and from all the solutions found for each $\epsilon$ we select the solution with minimum residual.
\textcolor{black}{
\subsubsection*{SINDy}
In their recent contribution, Brunton, Proctor and Kutz introduced SINDy (Sparse Identification of Nonlinear Dynamics) as a way to perform nonlinear system identification~\cite{Brunton2015}. Their main idea is, after formulating the inverse problem~\eqref{eq:inverse_rep}, to seek a {\it sparse} solution. In particular, given that Lasso can be computationally costly, they proposed to use sequential least squares with (hard) thresholding as an alternative. For a (prechosen) threshold $\lambda$, the method starts from a least squares solution and abandons all basis functions whose corresponding parameter in the solution has absolute value smaller than $\lambda$; then the same is repeated for the data matrix associated with the remaining basis functions, and so on and so forth, until no more basis function (and the corresponding parameter) is removed.
For fairness of comparison, we present results of SINDy according to the best threshold parameter $\lambda$ manually chosen so that no active basis function is removed in the very first step (see KSE example); for the Lorenz system example, we choose $\lambda=0.02$ as used in a similar example as in Ref.~\cite{Brunton2015}.}

\textcolor{black}{
\subsubsection*{Tran-Ward (TW)}
In their recent paper~\cite{TranWard} Tran and Ward considered the SID problem where certain fraction of data points are corrupted, and proposed a method to simultaneously identify these corrupted data and reconstruct the system assuming that the corrupted data occurs in sparse and isolated time intervals. In addition to an initial guess of the solution and corresponding residual, which can be assigned using standard least squares, the TW approach requires a pre-determiniation of three additional parameters: a tolerance value to set the stopping criterion, threshold value $\lambda$ used in each iteration to set those parameters whose absolute values are below $\lambda$ to be zero, and another parameter $\mu$ to control the extent to which data points that do not (approximately) satisfy the prescribed model are to be considered as ``corrupted data" and removed. For the Lorenz system example, we used the same parameters as in Ref.~\cite{TranWard} whereas for the KSE example, we fix $\mu=0.0125$ (the same used in Ref.~\cite{TranWard} and select $\lambda$ similarly as for the implementation of SINDy.}

\subsection*{Implementation Details of Entropic Regression (ER)}
As described in the main text, and as shown in details in Algorithm~(\ref{alg:ER}), a key quantity to compute in ER is the conditional mutual information $I(X;Y|Z)$ among three (possibly multivariate) random variables $X$, $Y$ and $Z$ via samples from these variables, denoted by $(x_t,y_t,z_t)_{t=1,\dots,T}$. Since the distribution of the variables and their dependences are generally unknown, we adopt a nonparametric estimator for $I(X;Y|Z)$ which is based on statistics of $k$ nearest neighbors~\cite{Kraskov2004}. We fix $k=2$ in all of the reported numerical experiments; we have found that the results change quite minimally when $k$ is varied from this fixed value, suggesting relative robustness of the method.

Another important issue in practice is the determination of threshold under which the conditional mutual information $I(X;Y|Z)$ should be regarded zero. In theory $I(X;Y|Z)$ is always nonnegative and equals zero if and only if $X$ and $Y$ are statistically independent given $Z$, but such absolute criterion needs to be softened in practice because the estimated value of $I(X;Y|Z)$ is generally nonzero even when $X$ and $Y$ are indeed independent given $Z$.
A common way to determine whether $I(X;Y|Z)=0$ or $I(X;Y|Z)>0$ is to compare the estimated value of $I(X;Y|Z)$ against some threshold. See Sec.~(\ref{Sec:SI_EntropicRegression}) for details of robust estimation of the threshold in the context of SID.

\section*{Supplementary Material}
See supplementary material for more details in information theory measures, and additional numerical results for the double-well potential, Lorenz system, and a coupled network of the logistic map.

\section*{Acknowledgements}
This work was funded in part by the Simons Foundation Grant No. 318812, the Army Research Office Grant No. W911NF-16-1-0081, the Office of Naval Research Grant No. N00014-15-1-2093, and also DARPA.

\bibliographystyle{plain}
\bibliography{PE_Refrences}




%
\newpage
\setcounter{section}{0}
\setcounter{equation}{0}
\setcounter{figure}{0}
\setcounter{page}{1}
\renewcommand{\thesection}{SI.\arabic{section}}
\renewcommand{\thesubsection}{\thesection.\arabic{subsection}}
\renewcommand{\thepage}{SI.\arabic{page}}
\renewcommand\thefigure{SI.\arabic{figure}}
\renewcommand\theequation{SI.\arabic{equation}}
\newpage

\title{How Entropic Regression Beats the Outliers Problem \\in Nonlinear System Identification\\ \textit{Supplementary Information (SI)}}
\maketitle


\tableofcontents

\section{Governing Dynamics, Over-sparsity, and Sensitivity for Expansion Order}
In the main text we briefly discussed the effects of polynomial expansion order choses to construct the basis matrix $\Phi$, and in this section we provide results from numerical experiments to supplement these claims.

Recall from our main text Eq.(26), in Compressive Sensing (CS) framework we solves the constrained optimization problem:
\begin{equation} \label{eq:CS_SI}
\begin{cases}
\argmin_{\bm{a}}\|\bm{a}\|_1,\\
\mbox{subject to}~\|\Phi\bm{a}-\bm{f}\|\leq\epsilon,
\end{cases}
\end{equation}
where the parameter $\epsilon\geq0$ is used to relax the otherwise strict constraint $\Phi\bm{a}=\bm{f}$, to allow for the presence of noise in data.

\begin{figure}[hbt]
\centering
\includegraphics[scale=1]{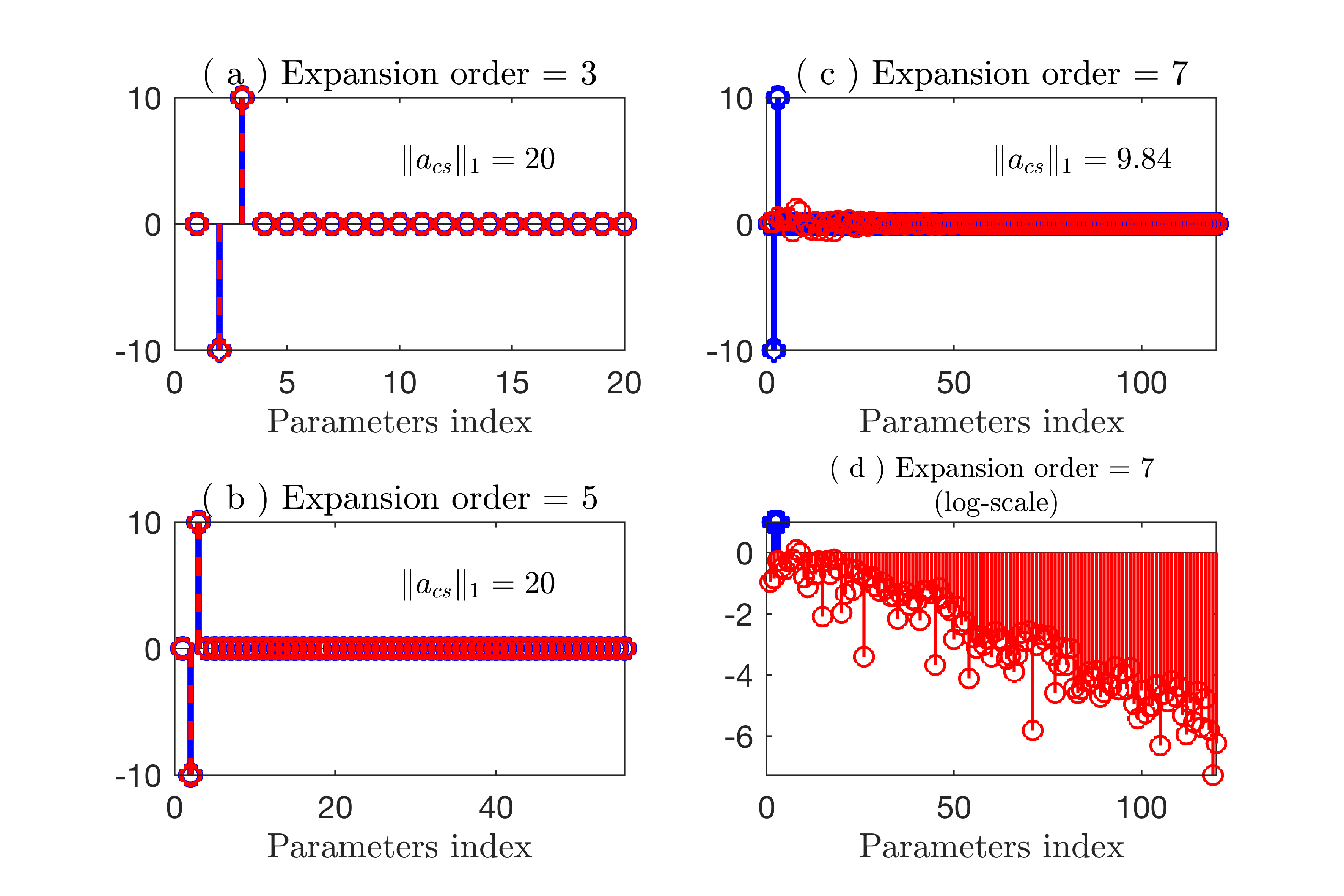}
\caption{CS reconstructed model for the first equation of Lorenz system regarding $\dot{x}$. The Solution shown in $\log_{10}$-scale in the $y$-axis for the parameters magnitude. From left to right, we see the recovered solution using the $3^{rd}$, $5^{th}$ and $7^{th}$ expansion order respectively. We used 300 noise free measurement, ($\epsilon_1 = \epsilon_2 = 0$). We see that with the $3^{rd}$ order polynomial expansion, CS recovered the solution with high accuracy, and it the same case with $5^{th}$ order polynomial expansion although the accuracy is slightly reduced, but we can still see the accurate sparse structure clearly. With the $7^{th}$ order polynomial expansion which produce 120 basis, we see a complete failure of CS where it over sparse the solution to have $\Vert \bm{a}_{cs} \Vert_{1} = 0.005$.}
\label{fig:CS357}
\end{figure}

\begin{figure}[hbt]
\centering
\includegraphics[scale=1]{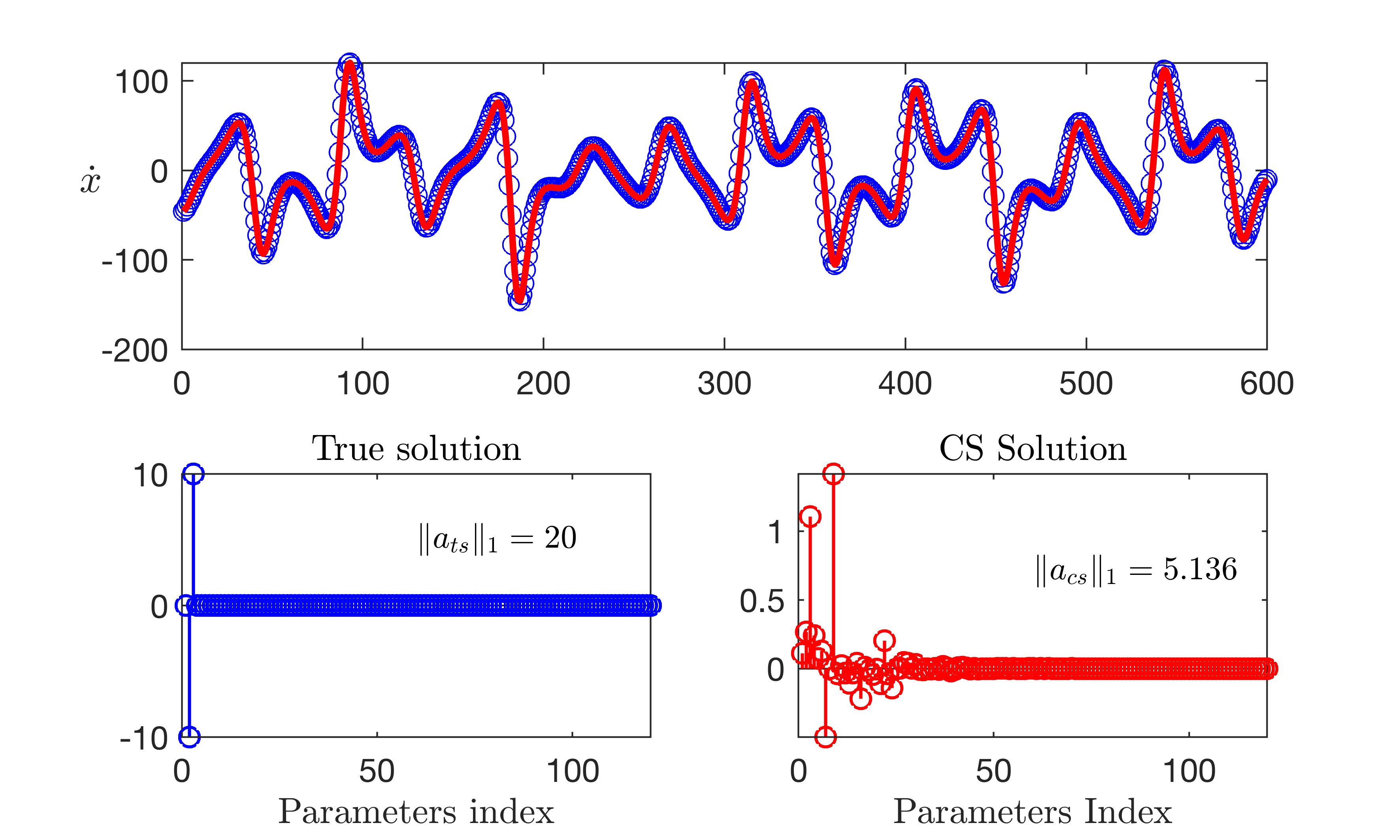}
\caption{CS Recovered solution for the first term $\dot{x}$ of Lorenz system using 600 noise free measurement, ($\epsilon_1 = \epsilon_2 = 0$). We see that even when we doubled the measurements, the CS is still over-sparse, although we have a good fitting curve, but the recovered system is wrong. In the other hand, the CS performs poor in recovering such dynamic with the presence of noise even with considering a low expansion order. \href{https://www.youtube.com/watch?v=CzsBS8wx6b8}{Click here} for a simulation of CS results of the same current example with considering the $3^{rd}$ order polynomial expansion and the presence of noise.}
\label{fig:CS7}
\end{figure}

Fig.~\ref{fig:CS357} shows the reconstructed model by CS for the first equation of Lorenz system regarding $\dot{x}$. We observe sensitivity on expansion order and how CS yields over-sparse solution at the $7^{th}$ order expansion. This result was obtained using 300 measurements. To extend the investigation we repeat the same numerical experiment with doubled number of measurements, and Fig.~\ref{fig:CS7} shows that CS still over-sparse the solution. This shows the relative sensitivity of CS with respect to the expansion order of basis functions.

\begin{figure}[t]
\centering
\includegraphics[scale=1]{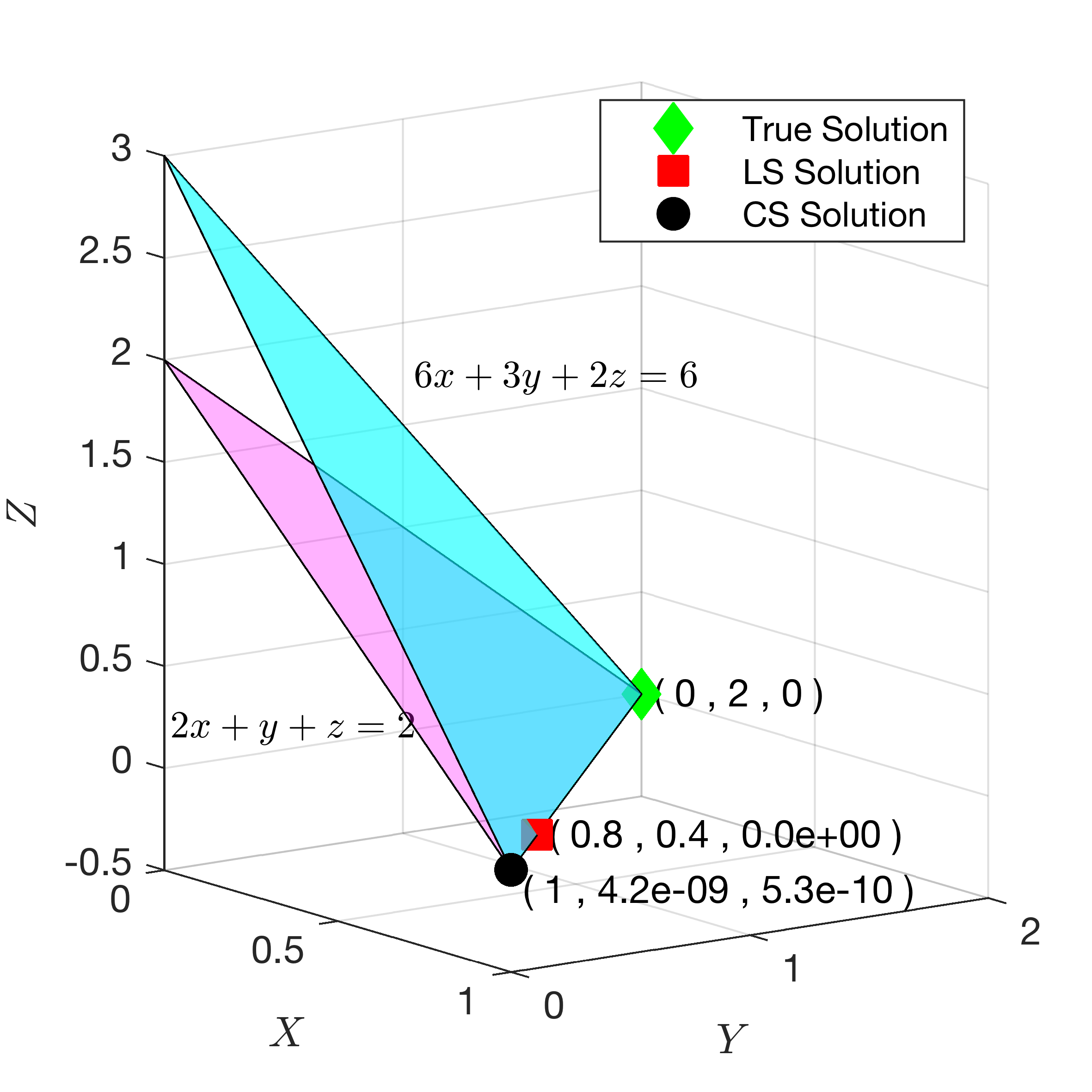}
\caption{Oversparsity: The line of intersection of the two planes (triangles) shows the solution plane. We see that compressed sensing solution is oversparsed.}
\label{fig:oversparsed}
\end{figure}

In order to construct a second example that clearly shows the oversparse mechanism in CS, consider the three-dimensional linear system:
\begin{eqnarray}
\left( \begin{array}{ccc} 6 & 3 & 2 \\ 2 & 1 & 1 \\ 1 & 2 & 1 \end{array} \right) \bm{a} = \left( \begin{array}{c} 6 \\ 2 \\ 4 \end{array} \right).
\end{eqnarray}
It is easy to find that the solution for the above system is $\bm{a} = \left( \begin{array}{ccc} 0 & 2 & 0 \end{array} \right)^{T}$. Now, suppose that the third ``measurement'' is missing, and we have the under-determined system
\begin{eqnarray} \label{eq:csexample}
\left( \begin{array}{ccc} 6 & 3 & 2 \\ 2 & 1 & 1 \end{array} \right) \bm{a} = \left( \begin{array}{c} 6 \\ 2 \end{array} \right)
\end{eqnarray}
then we have infinitely many solutions lies on the line of intersection of the two planes:
\begin{equation}
\begin{cases}
6x + 3y + 2z &= 6 \nonumber \\
2x + y + z   &= 2  \nonumber
\end{cases}
\end{equation}
Figure \ref{fig:oversparsed} shows this simple example, where the solution for $\bm{a}$ lies on the intersection of the two planes shown, and we see the true solution, the LS solution and CS solution on the solution line. We see how the least square solution is far from the true solution with a high margin of error, but we also see that it only invest in $x$ and $y$ direction where the line of intersections of the two plans lies, then, LS ignore the $z$ direction and try to invest in all feasible directions to reach the best residual.\\

CS have different mechanism, since within all feasible solutions, it tends to select the one with minimum  $\Vert \bm{a} \Vert_{1}$, even if there is another solution with the same number of sparse that have a residual $\Vert A\bm{a}-b \Vert_{2} = 0$, and it is the case in our example where $\Vert A\left( \begin{array}{ccc} 0 & 2 & 0 \end{array} \right)^{T} - b \Vert_{2} = 0$, while the CS solution has the residual $\Vert A \bm{a}_{CS}- b \Vert_{2} = 2.5\times 10^{-5}$. In other words, for the system $A\bm{a} = b$, if there exist two solutions such that $\Vert \bm{a_1} \Vert_{1} < \Vert \bm{a_2} \Vert_{1}$ and $\Vert A\bm{a_2}-b \Vert_{2} < \Vert A\bm{a_1}-b \Vert_{2} \leq \epsilon$, where $\epsilon$ is the tolerance for CS optimization, then CS will select $a_1$ as a solution, even it has higher residual, and regardless of the structure of the sparse or the information flow between the basis functions and the observations. Numerically, assume the system in Eq.~\ref{eq:csexample} to be:
\begin{eqnarray}
\left( \begin{array}{ccc} (6+1e^{-10}) & 3 & 2 \\ 2 & (1+1e^{-16}) & 1 \end{array} \right) \bm{a} = \left( \begin{array}{c} 6 \\ 2 \end{array} \right)
\end{eqnarray}
and consider a reasonable tolerance for CS solver to be $\epsilon = 1e^{-9}$, then CS will always pick [1 0 0] as a solution even it have higher residual.

For many applications..., it is accepted to have such solution since it lies on the solution line and such residual difference will have negligible effect on the final result, But in discovering the governing equations of dynamical systems, such solution can often lead to a completely wrong structure of the system.

\section{Information Theory}
\textcolor{black}{In this section we review some basic concepts in information theory that underlie the development of entropic regression. These include classical concepts such as entropy and mutual information as well as recent developments of transfer entropy and causation entropy.}


\subsection{Entropy}
Entropy is firstly known as an extensive property of a thermodynamic system. The entropy of a thermodynamic system is a function of the microscopic states consistent with the macroscopic quantities that characterize the system. Assuming equal probability of the microscopic states, the entropy is given by:
\begin{equation} \label{eq:Boltz}
S = k_{B} \ln(W)
\end{equation}
where $W$ is the number of microscopic states and $k_{B}$ is Boltzmann constant named after Ludwig Eduard Boltzmann where the Eq.~\ref{eq:Boltz} curved on his gravestone. Boltzmann saw entropy as a measure of statistical disorder in the system.

\begin{figure}[ht]
\centering
\includegraphics[scale=1]{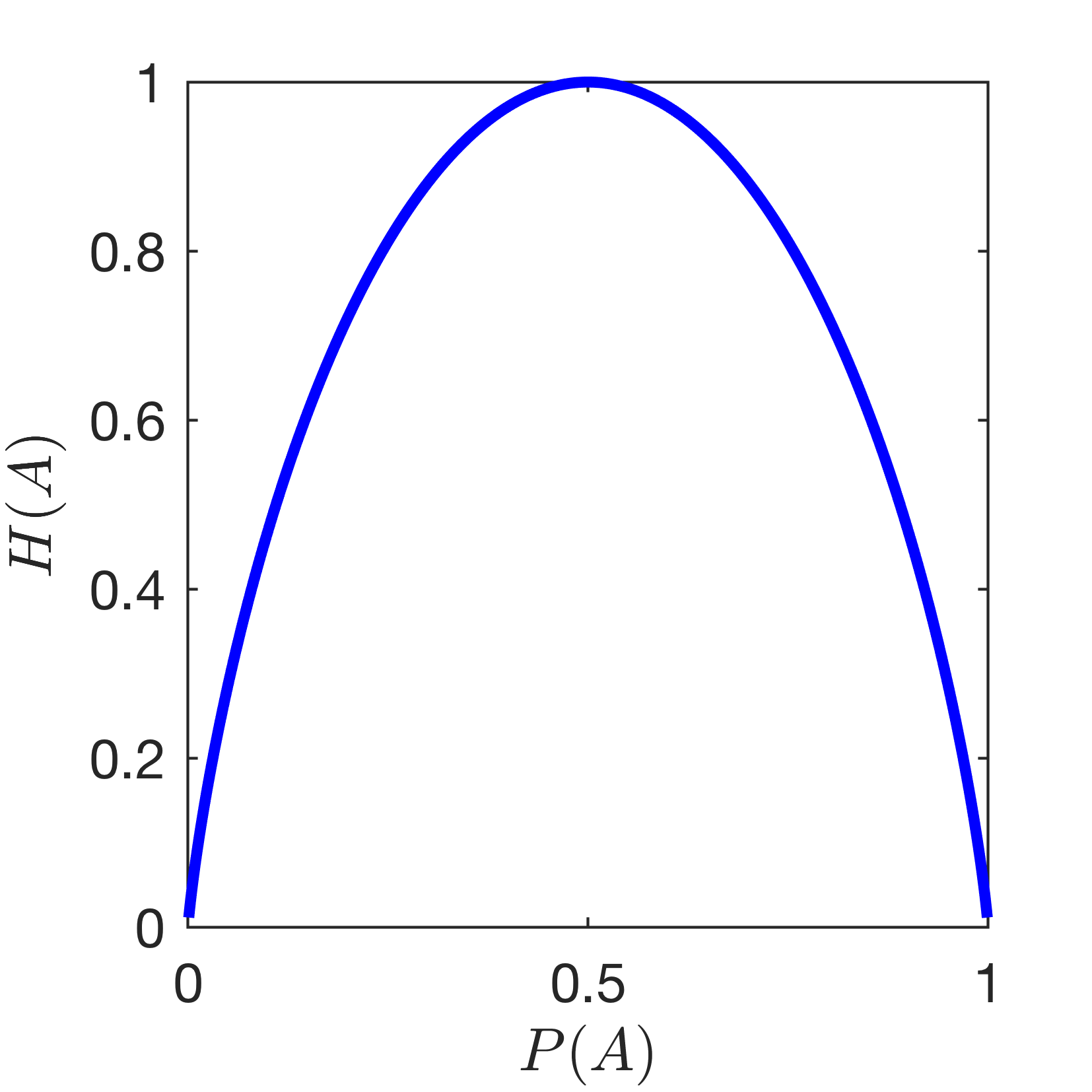}
\caption{Entropy of the event $A$. Here we assume the states to be the occurrence and non-occurrence of the event $A$, and $P(A)$ represent the probability of the occurrence state. This figure show the uncertainty about the event $A$ occurrence. In $x$-axis we have the probability $P(A)=p$ that the event $A$ occurs, then by Eq.~\ref{eqn:Entropy} and considering the $\log$ to base 2, $H(A) = -p\log(p) - (1-p)\log(1-p)$ is the measure of uncertainty of the event $A$, where $(1-p)$ is the probability that the event $A$ will not occur. Starting from $P(A) = 1$, meaning that the event $A$ is always occurs or it is the only event we have, then $H(A) = 0$, meaning that there is no uncertainty and we are sure of the event $A$ occurrence. As the probability decrease, the entropy (uncertainty) increase to reach its maximum at $P(A) = 0.5$. Continuing decreasing $P(A)$ will reduce the entropy again, since we become more certain that the event $A$ will not occur, until we become completely certain that $A$ will not occur with $H(A)=0$ at $P(A)=0$.}
\label{fig:EntFig}
\end{figure}

An analog to thermodynamic entropy is information entropy introduced by Claude Shannon in 1948 as ``measures of information, choice, and uncertainty''. To describe Shannon's entropy, consider a discrete random variable $X$ whose probability mass function is denoted by $p(x) = Prob(X = x)$. One can calculate its entropy as \cite{Cover2005, Shannon1948},
\begin{equation} \label{eqn:Entropy}
H(X) = -K\sum_{x} p(x) \log p(x),
\end{equation}
where $K$ is positive constant, and $H(X)$ is a measure of the uncertainty or unpredictability of $X$. Note that if we assume uniform probability distribution for the states of $X$, then we have $p(x) = \frac{1}{N}$, where $N$ is the number of states, and then Eq.~\ref{eqn:Entropy} can be written as $H(X) = K\log(N)$ similar to Boltzmann's entropy under the same assumption of equal probability of the states. The constant $K$, as Shannon sates, is merely amounts to a choice of a unit of measurement, and we consider $K = 1$ for the rest of this document for simplicity. Fig.~\ref{fig:EntFig} shows the entropy function for a random event with different probability.

Shannon's work provided extended and generalized view and understanding for the entropy, and one of the extended perspectives of Shannon's entropy is dealing with the continuous random variables, and it takes the form:
\begin{equation}\label{eq:EntCont}
H(X) = \int_{-\infty}^{\infty} f_{X}(x)\log(f_{X}(x)) dx,
\end{equation}
where $f_{X}(x)$ is the probability density function. The entropy shown in Eq.~(\ref{eq:EntCont}) is referred to the \textit{differential entropy}.

\subsection{Mutual Information}
The entropy defined in Eq.~(\ref{eqn:Entropy}) naturally extends to the case of multiple random variables. For example, the joint entropy $H (X , Y )$, and conditional entropy $H (X | Y )$ of two random variables $X$ and $Y$ is given, respectively, by \cite{Cover2005, Shannon1948}, 
\begin{equation}
H(X,Y) = -\sum_{x,y} p(x,y) \log p(x,y)
\end{equation}
\begin{eqnarray}
H (X | Y ) &=& -\sum_{y} p(y) H(Y | X=x) \notag \\
                            &=& -\sum_{x,y} p(x,y) \log p(x | y),
\end{eqnarray}
where $p(x,y)$ is the joint probability distribution, and  $H (X | Y )$ (read as entropy of $X$ given $Y$) is the measure of the uncertainty in $X$ if $Y$ is known. Some of the main properties of the entropy, joint entropy, and conditional entropy can be summarize as follows:
\begin{itemize}
\item The entropy of a discrete variable $X$ is positive ($H(X) \geq 0$), while the differential entropy does not satisfy this property.
\item For two independent random variables $X$ and $Y$, $H(X,Y) = H(X) + H(Y) $.
\item The chain rule: $H(X,Y) = H(X) + H(Y|X)$.
\item One important property is that for a random variable $X$, the conditional entropy of $X$ given any other variable $Y$ will reduce the entropy of $X$, meaning that $H(X) \geq H(X|Y)$. The equality holds when $X$ and $Y$ are independent with $H(X,Y) = 0$. This property tells that the information comes from $Y$ reduces the uncertainty about $X$, and when $Y = X$, that means we have given all the information about $X$, and then we become completely certain about $X$, and that gives $H(X|X) = 0$.
\end{itemize}

The joint and conditional entropies can lead to a measures that detect the statistical dependence or independence between random variables. Such measure is called the mutual information between $X$ and $Y$, and it is given by \cite{Cover2005, Shannon1948},
\begin{eqnarray}
I (X ; Y ) &=& H (X ) - H (X |Y )   \notag \\
             &=& H (Y ) - H (Y |X ) \notag \\
             &=& H(X) + H(Y) - H(X,Y),
\end{eqnarray}
where the mutual information $I (X ; Y )$ (reads as mutual information between $X$ and $Y$) is a measure of the mutual dependence between the two variables. In terms of joint probability distribution, mutual information can be written as,
\begin{equation}\label{eq:midisc}
I(X;Y) = \sum_{y \in Y} \sum_{x \in X} p(x,y)\log \left( \frac{p(x,y)}{p(x)p(y)} \right),
\end{equation}
and in its continuous form,
\begin{equation}\label{eq:micont}
I(X;Y) = \int_{Y} \int_{X} f_{X,Y}(x,y)\log \left( \frac{f_{X,Y}(x,y)}{f_{X}(x)f_{Y}(y)} \right),
\end{equation}
where $f_{X,Y}(x,y)$ is the joint probability density function for the two continuous random variables $X$ and $Y$.

In case of independence of the two random variables, we have
\begin{equation}
p(x,y) = p(x)p(y),
\end{equation}
and then we have 
\begin{equation}
\log \left( \frac{p(x,y)}{p(x)p(y)} \right) = \log (1) = 0 \implies I(X;Y) = 0.
\end{equation}
The same principle holds for the continuous variables in Eq.~(\ref{eq:micont}), while $I(X;Y)$ satisfy the inequality $I(X;Y) \leqslant min[H(X),H(Y)]$ only in the discrete variables case.

\subsection{Transfer Entropy and Causation Entropy}

For two stochastic processes $X_{t}$ and $Y_{t}$, the reduction of uncertainty about $X_{t+1}$ due to the information of the past $\tau_{Y}$ states of $Y$ , represented by 
$$Y^{(\tau_{Y})}=(Y_{t},Y_{t-1} ,...,Y_{t-\tau_{Y} + 1}),$$
in addition to the information of the past $\tau_{X}$ states of $X$ , represented by
$$X^{(\tau_{X})}=(X_{t},X_{t-1} ,...,X_{t-\tau_{X} + 1}),$$
this reduction of uncertainty about $X_{t+1}$ is measured by ``Transfer Entropy'' which given by \cite{Cover2005, Bollt2014},
\begin{equation}
T_{Y \rightarrow X} = H(X_{t+1}|X{t}^{\tau_{X}}) -  H(X_{t+1}|X{t}^{\tau_{X}},Y_{t}^{\tau_{Y}}).
\end{equation}

The traditional approach of inferring causality between two stochastic processes is to perform the Granger causality test \cite{Granger1969}. The main limitation of this test is that it can only provide information about linear dependence between two processes, and therefore fails to capture intrinsic nonlinearities that are common in real-world systems. To overcome this difficulty, Schreiber developed the concept of transfer entropy between two processes \cite{Schreiber2000}. Transfer entropy measures the uncertainty reduction in inferring the future state of a process by learning the (current and past) states of another process.

In our work \cite{Bollt2014,Sun2014a}, we showed by several examples that causal relationship inferred by transfer entropy are often misleading when the underlying system contains indirect connections, a dominance of neighboring dynamics, or anticipatory couplings. For example, referring to the main text and the equation $\bm{f} = \Phi \bm{a}$, we see that the approaches that consider the transfer entropy in order to find the weak terms in $\Phi$ that has no influence on $\dot{X}$ to construct the sparse matrix $\bm{a}$, these approaches neglect a simple and clear idea that the terms of $\Phi$ has an indirect influence on $\bm{f}$ through the other terms of $\Phi$. To account for these effects, we developed a measure called \textit{Causation Entropy} (CSE) \cite{Bollt2014,Sun2014a}, and show that its appropriate application reveals true coupling structures of the underlying dynamics.

Consider a stochastic network of $N$ processes (nodes) denoted by:
\begin{equation}
X_{t} = \{ X_{t}^{(1)}, X_{t}^{(2)}, \dots, X_{t}^{(N)} \}
\end{equation}
where $X_{t}^{(i)} \in \mathbb{R}^{d}$ is a random variable representing the state of process (or node) $i$ at time $t$, and $i \in \mathcal{V} = \{1,2,\dots, N\}$, and let $I,J$, and $K$ be a subsets of $\mathcal{V}$, then we can define the causation entropy as the following:\\

\noindent \textbf{Definition 1} \cite{Sun2014a}: The causation entropy from the set of processes $J$ to the set of processes $I$ conditioning on the set of processes $K$ is defined as
\begin{equation}
C_{ J \rightarrow I | K } = H(X_{t+1}^{(I)}|X_{t}^{(K)}) -  H(X_{t+1}^{(I)}|X_{t}^{(K)},X_{t}^{(J)}).
\label{eq:CSE}
\end{equation}

The Causation entropy is a natural generalization of transfer entropy from measuring pairwise causal relationships to network relationships of many variables. In particular, we can list the main properties for the causation entropy, noting that if $J = \{j\}$ and $I = \{i\}$, we simplify the notation as $C_{ j \rightarrow i | K }$:
\begin{itemize}
\item If $j \in K$, then the causation entropy $C_{ j \rightarrow i | K } = 0$, as $j$ does not carry extra information (compared to that of $K$).
\item If $K = \{ i \} $, then the causation entropy recovers the transfer entropy  $C_{ j \rightarrow i | i } = T_{ j \rightarrow i }$ which is given by $T_{ j \rightarrow i } = H(X_{t+1}^{(i)}|X_{t}^{(i)}) -  H(X_{t+1}^{(i)}|X_{t}^{(i)},X_{t}^{(j)})..$
\end{itemize}

In \cite{Sun2014a}, we introduced the principle of optimal Causation Entropy (oCSE) in a network of $N$ processes to find the minimum subset that maximizes the causation entropy. This minimal subset can be seen as the dominant subset of a network of $N$ processes, and they rule the underlying dynamic of the network. and in the same principle, we are looking for the dominant terms of the basis function $\Phi$ on the system dynamic $\bm{f}$. See (main text Fig.1) for visualization of the reformulation of Lorenz system in a network of processes.

%
%
\section{Entropic Regression} \label{Sec:SI_EntropicRegression}
In our main text we discussed the Entropic Regression method and provided its Algorithm (main text Algorithm 1). In this section we discuss the tolerance estimation and its effect on the performance of ER.

In our previous work \cite{Sun2014a} we introduced a standard shuffle test, with a ``confidence'' parameter $\alpha\in[0,1]$ for tolerance estimation. The shuffle test requires randomly shuffling of one of the time series $n_s$ times, to build a test statistic. In particular, for the $i$-th random shuffle, a random permutation $\pi^{(i)}:[T]\rightarrow[T]$ is generated to shuffle one of the time series, say, $(y_t)$, which produces a new time series $(\tilde{y}^{(i)}_t)$ where $\tilde{y}^{(i)}_t=y_{\pi^{(i)}(t)}$; $(x_t)$ is kept the same. Then, we estimate the mutual information $I(X;\tilde{Y}^{(i)})$ using the (partially) permuted time series $(x_t,\tilde{y}^{(i)}_t)$, for each $i=1,\dots,n_s$. For given $\alpha$, we then compute a threshold value $I_\alpha(X;Y)$ as the $\alpha$-percentile from the values of $I(X;\tilde{Y}^{(i)})$. If $I(X;Y)>I_\alpha(X;Y)$, we determine $X$ and $Y$ as dependent; otherwise independent. 

We showed in \cite{Sun2014a}, the robustness of shuffling test for optimal causation entropy calculations specially in complex dynamics, although it is computationally expensive. For more efficient computations complexity, we considered two different approach for tolerance estimation with the confidence parameter $\alpha$; the Dynamic, and the Static approaches. 

In the \textbf{Dynamic} tolerance estimation, we start the forward step procedure (See main text Algorithm~(\ref{alg:ER})) with initial tolerance $tol = 0$, and we update the tolerance value at the end of the forward step procedure by the shuffle test shown in Algorithm~(\ref{alg:shflTest}) below. Given the confidence parameter $\alpha$, we update the tolerance to be the outcome of the shuffle test on the conditional mutual information of the forward step $I(\Phi R(\Phi,\bm{f},\{S_f,j\});\bm{f} | \Phi R(\Phi,\bm{f},\{S_f\}))$, with $j$ indicates the index of maximum mutual information found by the current forward step.

The \textbf{Static} approach is more computationally efficient,  where we apply the shuffle test to the mutual information $I(\bm{f};\bm{f})$ with the confidence parameter $\alpha$, and we assign the outcome of the shuffle test to the tolerance at the beginning of the forward step with no updates follows.

\begin{algorithm}
\caption{Shuffle Test}\label{alg:shflTest}
\begin{algorithmic}[1]
\Procedure{Shuffle Test}{$\bf{f},\Phi, S_f, j, \alpha, n_s $}
\State $i = 1$, $I = \emptyset$
\While{$i \leq n_s$}
\State  $\hat{I} \leftarrow I(\Phi R(\Phi,\bm{f},\{S_f,j\});\bm{f}_{\pi^{i}} | \Phi R(\Phi,\bm{f},\{S_f\}))$,
\State $i := i + 1$,
\EndWhile
\State \textbf{return} $\hat{I}$
\State $\mathcal{I} \leftarrow \hat{I}$ s.t. $\mathcal{I}_{j}\leq \mathcal{I}_{j+1}$, $j = 1,\dots,n_s-1$
\State $tol = \mathcal{I}_{k}$, where $k = \lceil \alpha n_s \rceil$.
\EndProcedure
\State \textbf{return} $tol$
\end{algorithmic}
\end{algorithm}

\section{Additional Numerical Results}
To demonstrate the utility of ER for nonlinear system identification under noisy observations, we compare its performance against existing methods including the standard least squares (LS), orthogonal least squares (OLS), Lasso, and compressed sensing (CS). The details of the existing approaches are described in the Methods Section. The examples we consider represent different types of systems and scenarios, including both ODEs and PDEs, differential and difference equations, and network-coupled dynamics. In addition, we consider different noise models and especially the presence of outliers in order to evaluate the robustness of the respective methods.

For each example system, we sample the state of each variable at a uniform rate of $\Delta{t}$ to obtain a multivariate time series $\{z_k(t_i)\}_{k=1,\dots,N;i=1,\dots,\ell}$; then we add noise to each data point and obtain the observed time series denoted by $\{\hat{z}_k(t_i)\}$, where
\begin{equation}
\hat{z}_k(t_i) = z_k(t_i) + \eta_{ki},
\end{equation}
with $\eta_{ki}$ represents noise.
%
%
\subsection{Double Well Potential} \label{sec:SI_doublewell}
In analogy to the example in our main text, we consider the equation
\begin{equation}\label{eq:doubleWell2}
f(x) = x^4 - x^2.
\end{equation}
and we sample 61 equally spaced measurements for $x \in [-1.2, 1.2]$, and we construct $\Phi$ using the $10^th$ order polynomial expansion with $K = 11$ is the number of candidate functions. Then, we consider a single fixed value corrupted measurement to be $f(0.52) = 0.5$.

\begin{figure}
\centering
\includegraphics[scale=1]{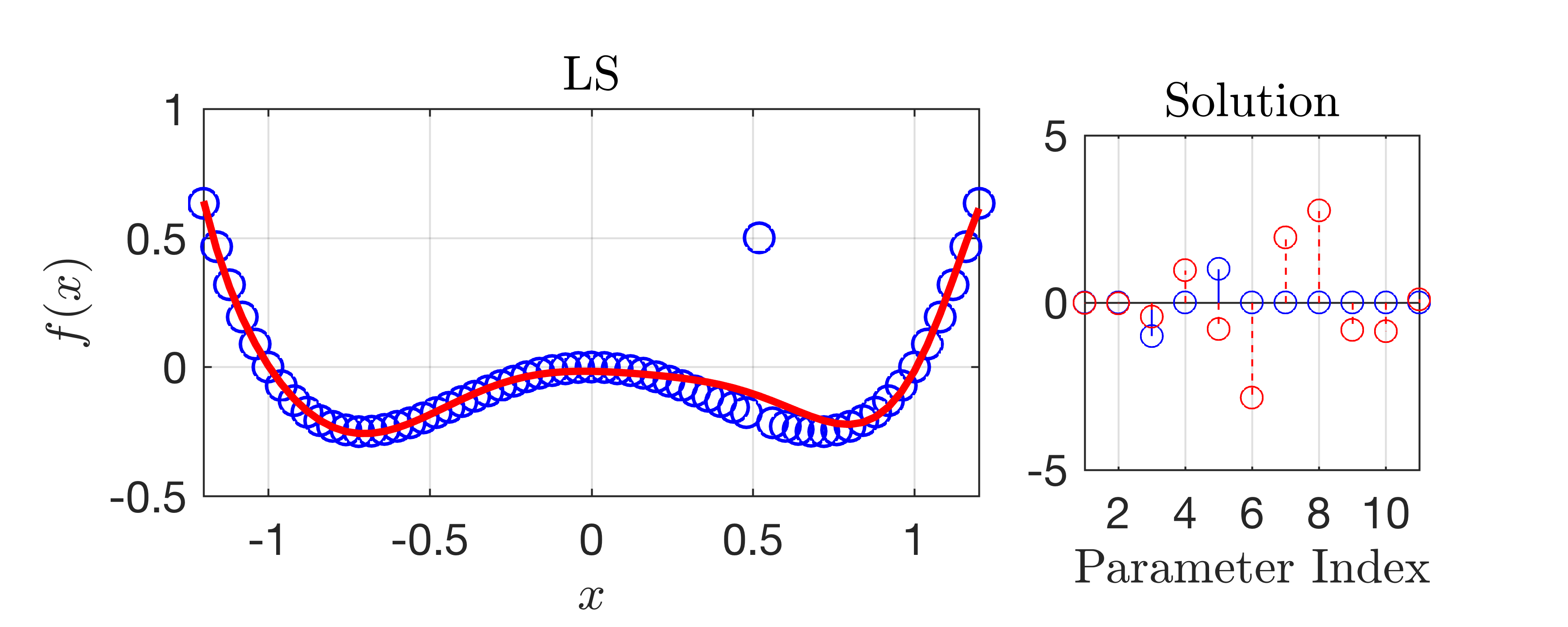}
\caption{The LS solution for the data given by Eq.~\ref{eq:doubleWell2}. This result shows how the LS invest in all available parameters to reach the best possible fitting. In fact, the residual of the least square solution was lower than the residual of the true solution, $0.6535 = \Vert \Phi \Phi^\dagger\bm{f} - \bm{f} \Vert < \Vert \Phi \bm{a_{true}} - \bm{f} \Vert = 0.6973$, and in sparse regression literature, this initiate the need for developing trade off algorithms that considers different measures such as $\Vert \bm{a} \Vert_{1}$ and $\Vert \bm{a} \Vert_{0}$. }
\label{fig:dwLS}
\end{figure}

\begin{figure}
\centering
\includegraphics[scale=1]{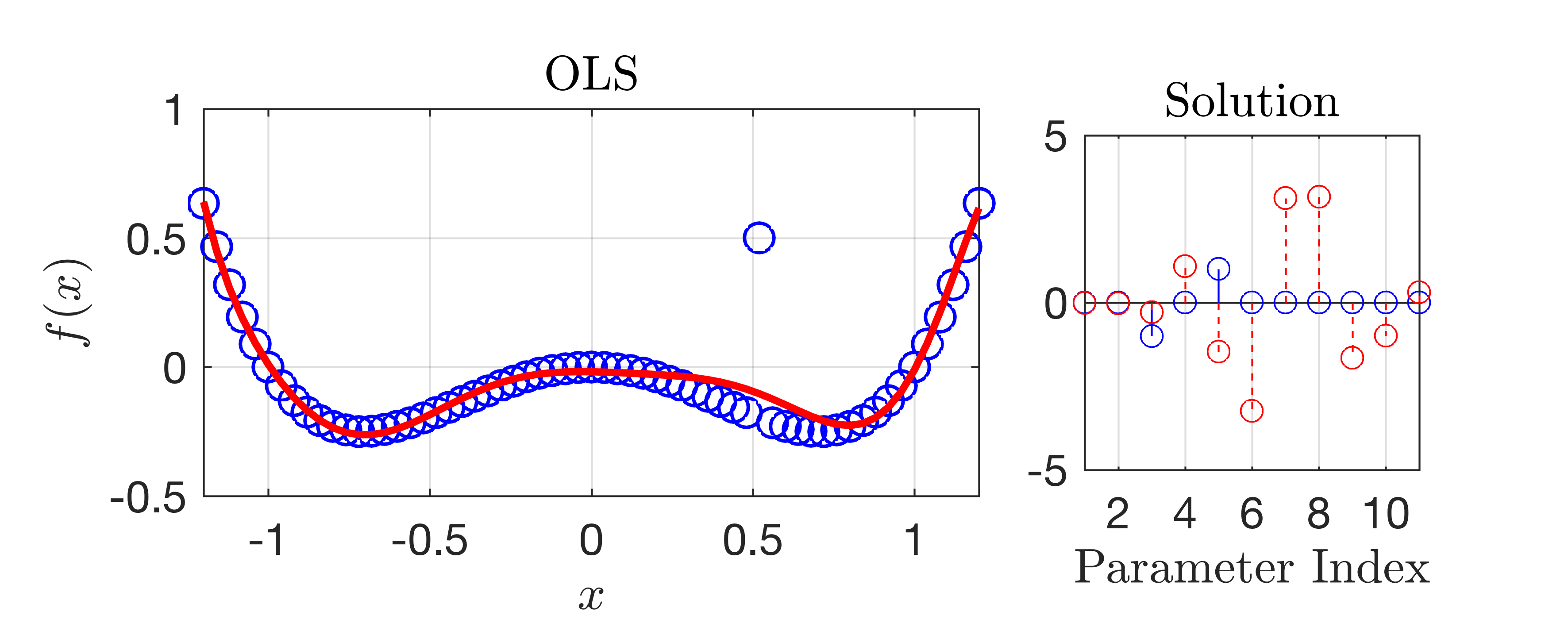}
\caption{The OLS solution with 1000 log-spaced span for the threshold value $\epsilon \in [10^{-6}, 10^{2}]$. We see that the OLS failed to find solution better than the LS and they are almost identical.}
\label{fig:dwOLS}
\end{figure}

In this example, we see that the true solution will have a residual $\delta$ equal to outliers deviation from its true position,
\begin{equation}
\delta = \sqrt[]{(f(0.52) - 0.5)^2} = 0.6973
\end{equation}
Fig.~\ref{fig:dwLS} shows the result for LS. The LS with its BLUE property (Best Linear Unbiased Estimator), succeed to minimize the residual to have better fitting residual than the true solution, but it is clear that the residual value does not reflect reliable solution. Practically, when the true solution gives a fitting residual $\delta$, then any other solution deviates in its residual from $\delta$ will have a reduction in the solution accuracy, no matter the direction of deviation from $\delta$. In Fig.~\ref{fig:dwOLS}, we see the result of OLS. We see that the results with the best residual of OLS is almost identical to LS result. Here it worth to say a detailed review for the 1000 OLS solutions under different threshold showed us a small interval that gives solutions closer in structure to the true solution more than the minimum residual solution is shown, which is if treated with suitable trade-off strategy can give a better solution. 

\begin{figure}
\centering
\includegraphics[scale=1]{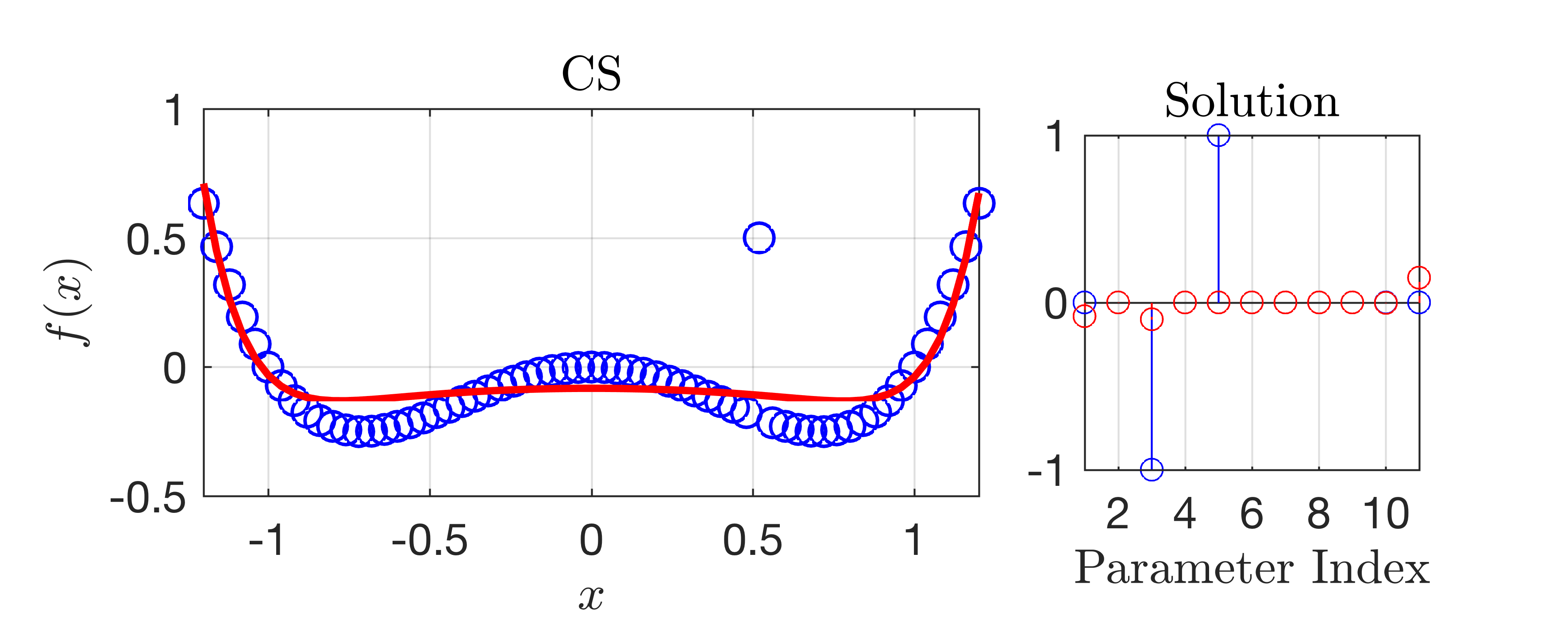}
\caption{The CS solution, with 1000 log-spaced span for $\epsilon \in [10^{-6}, 10^{2}]$. The solution with minimum residual is shown to the right. As expected, the CVX solver failed to find any feasible solution for all values of $\epsilon < 0.69$, and that was the reason to consider $10^{2}$ as the upper bound of epsilon although it represent a high value for tolerance.}
\label{fig:dwCS}
\end{figure}

\begin{figure}
\centering
\includegraphics[scale=1]{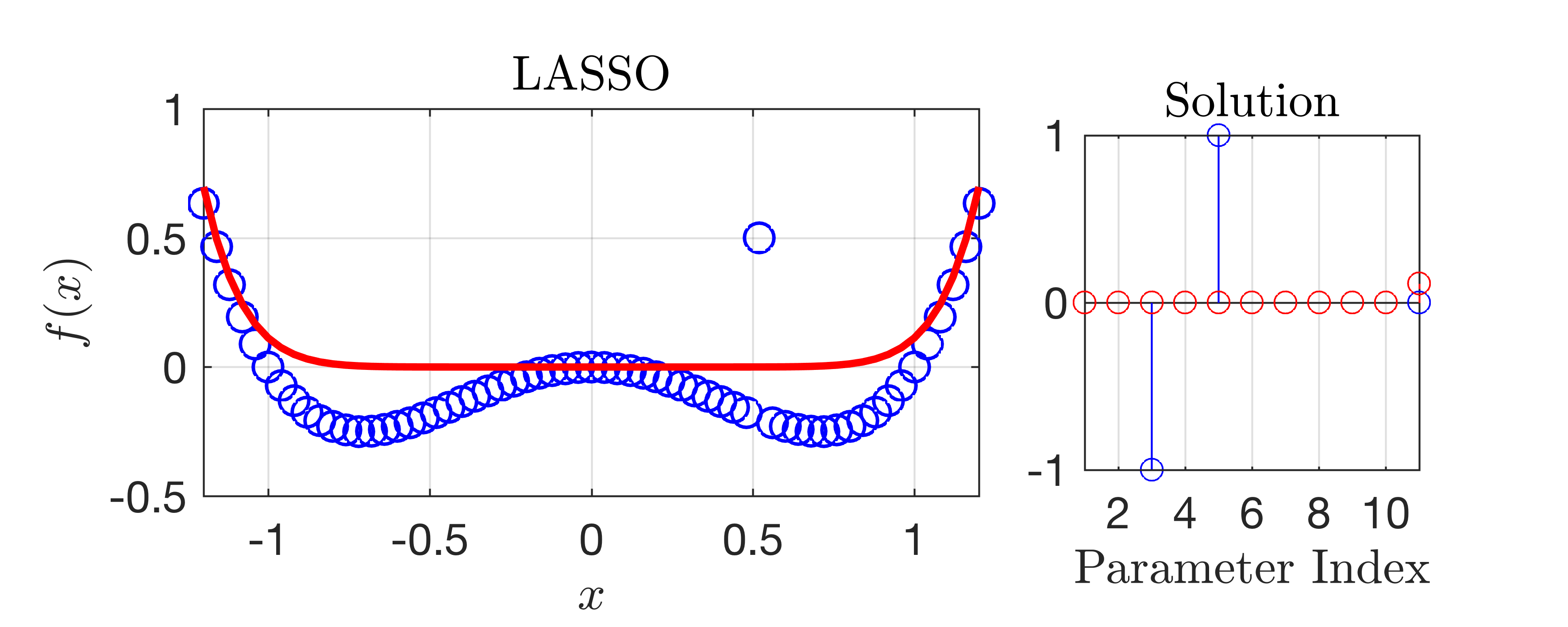}
\caption{The LASSO solution, with 1000 equally-spaced span for $\lambda \in \left[\Vert \Phi \Phi^\dagger\bm{f} - \bm{f} \Vert, \Vert \bm{f} \Vert\right]$. The solution with minimum residual is shown to the right and it found at $\lambda = 0.818$.}
\label{fig:dwLASSO}
\end{figure}

Fig.~\ref{fig:dwCS} shows the result for CS, where it failed to find any feasible solution for all values of $\epsilon < \delta$. Such outliers makes it hard to find a parameter vector $\bm{a}$ that can fit the data including the outliers point, and even with considering high resolution for $\epsilon$ span, so, CS as discussed before tends to select the solution with minimum $\Vert \bm{a} \Vert_{1}$ within the best feasible residuals. CS solution simulation for different outliers values is provided on our YouTube channel \href{https://www.youtube.com/watch?v=2FyfD9U0f6s&t=0s&index=3&list=PLKV2TLjSrnnoCQpbMYgD5neD1E-TGGfvE}{here}. Fig.~\ref{fig:dwLASSO} shows the result for LASSO, and it shows the sparse solution with wrong structure of LASSO. We considered the bounds of $\lambda$ to be $\lambda \in \left[\Vert \Phi \Phi^\dagger\bm{f} - \bm{f} \Vert, \Vert \bm{f} \Vert\right]$, where $\lambda = \Vert \Phi \Phi^\dagger\bm{f} - \bm{f} \Vert$ is the penalty on the solution with all entries are non-sparse and $\lambda = \Vert \bm{f} \Vert$ is the penalty on the solution with all entries are sparse. \textcolor{black}{For this example, different from others (see Methods section), and because of its small dimensions, we considered very large span (1000 values) of the tununing parameter value for OLS, LASSO and CS to investigate the best expected outcomes of the methods.}

\begin{figure}
\centering
\includegraphics[scale=1]{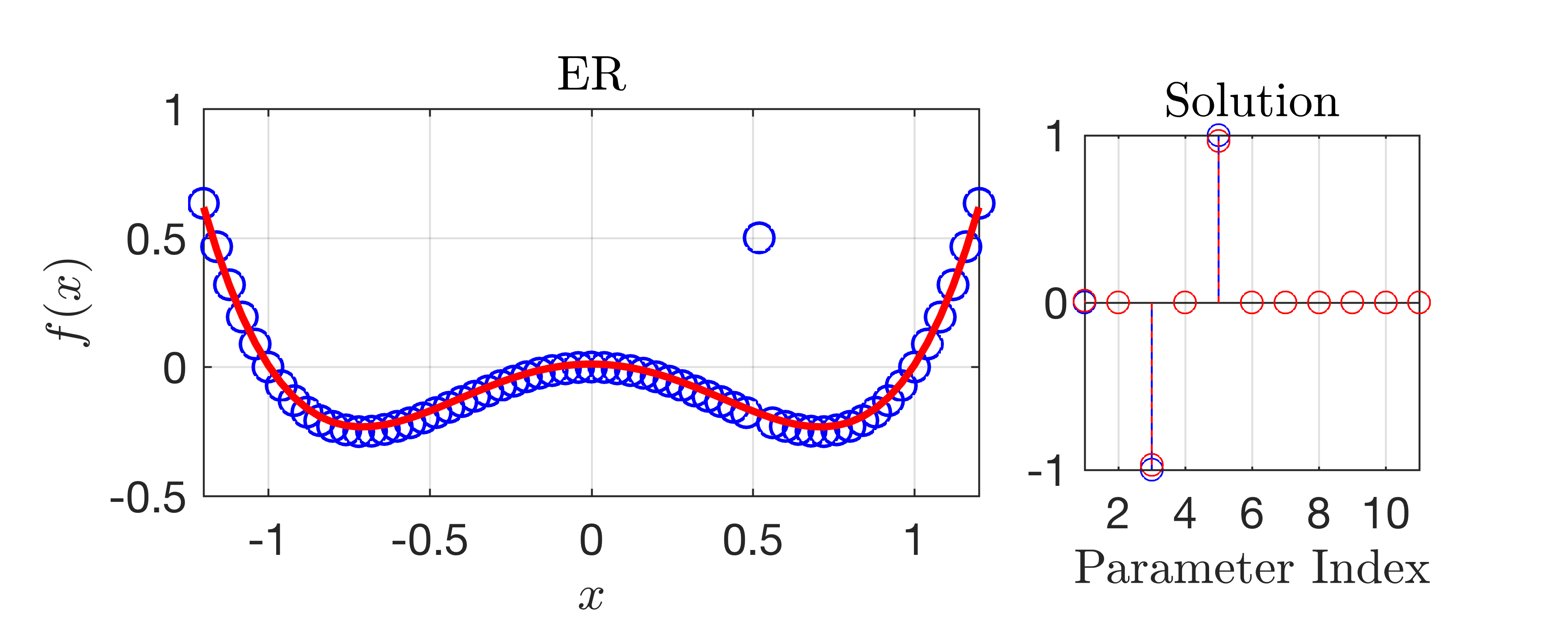}
\caption{The ER solution. We see that ER recovered the true solution, No trade-off, No-tuning parameter and large span with expensive computations.}
\label{fig:dwER}
\end{figure}

Fig.~\ref{fig:dwER} shows the accurate structure found by ER. Even with a slight difference in the magnitude of the parameters, we see how ER recovers the true basis functions. The residual of ER was $0.865$, which is higher more than most other methods, but the ER focuses on the information flow between the basis and dynamic and not the residual of solution magnitudes.

\begin{figure}
\centering
\includegraphics[scale=1]{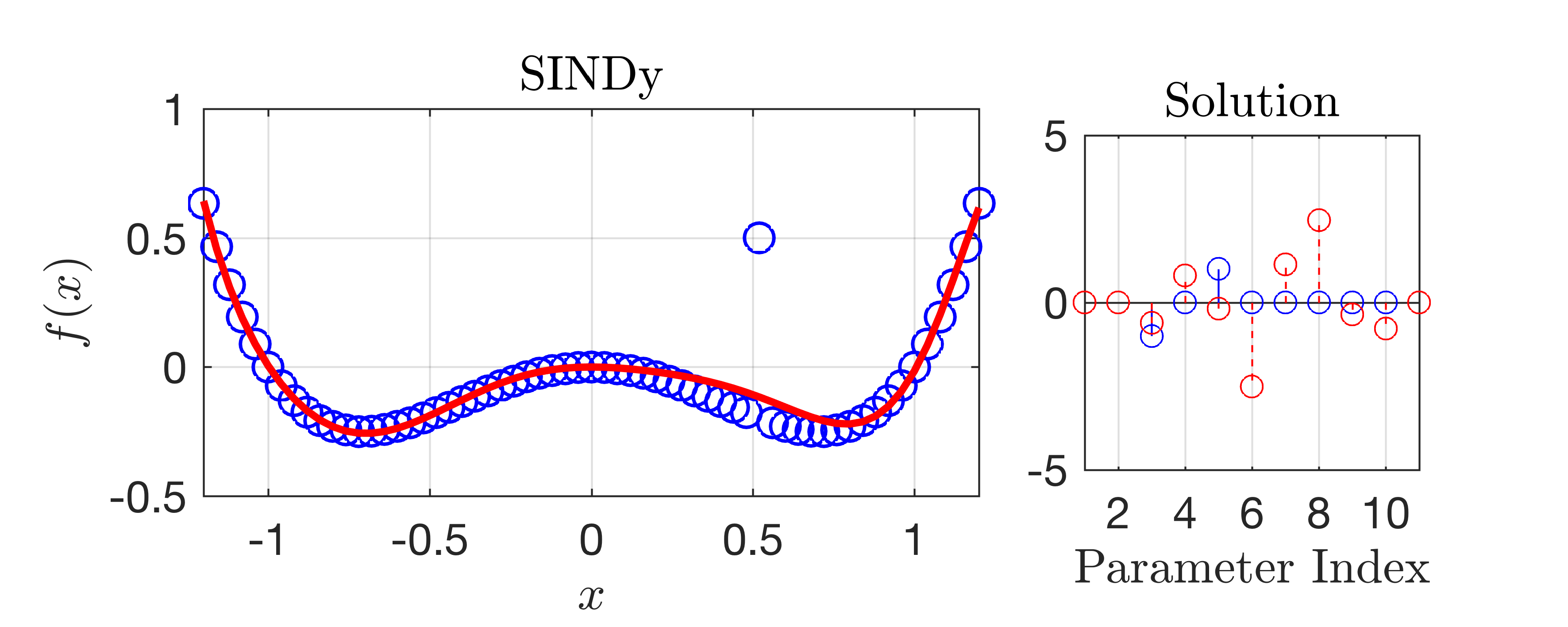}
\caption{SINDy solution. We choose the threshold value of SINDy to be $\lambda = 0.42$, which is the optimal value (chosen manually since there is no unsupervised method for such choice) that prevent SINDy from oversparse the true parameters.}
\label{fig:dwSINDy}
\end{figure}

\begin{figure}
\centering
\includegraphics[scale=1]{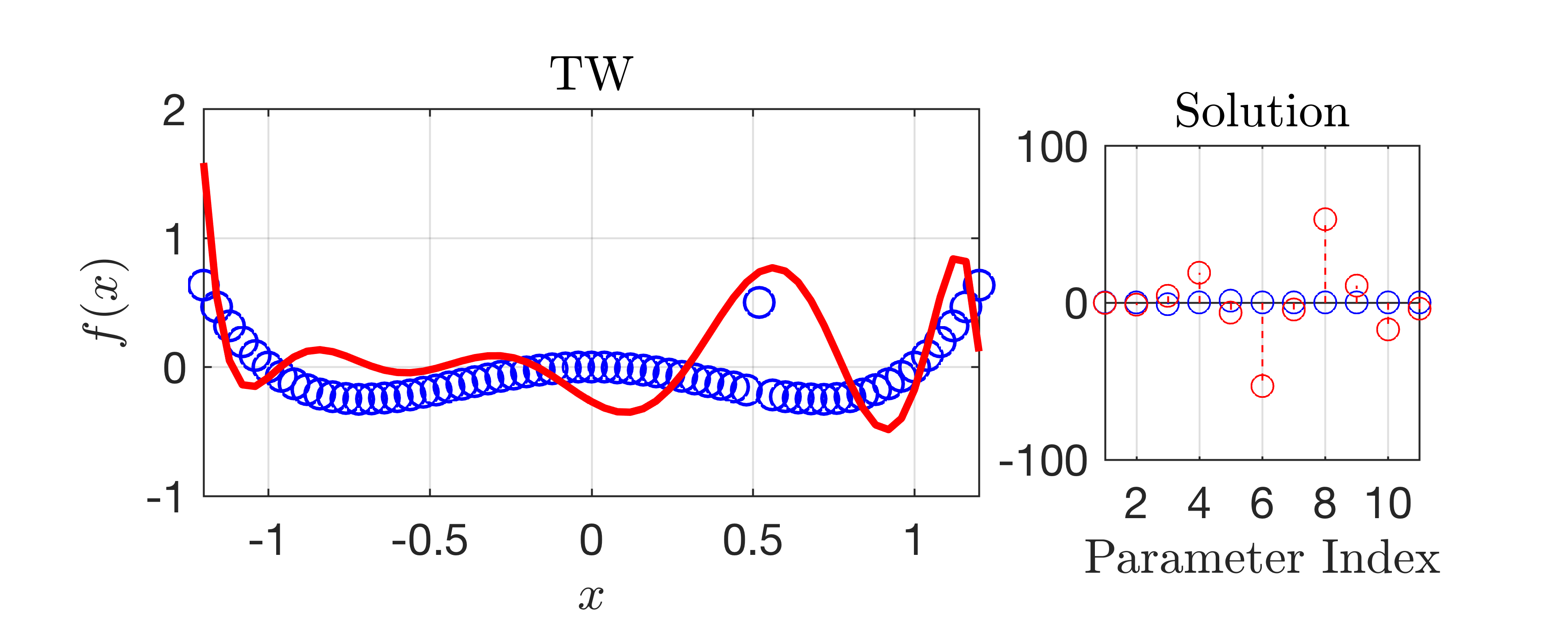}
\caption{TW solution. Under the default values for TW method, $\mu = 0.0125$ and $\lambda = 0.1$, the results was very poor, and that was surprising since the problem setting match the exact assumptions of availability of ``exact'' measurement, and here we assume only one outlier point. So, in analogy to Fig.~\ref{fig:TWcontourplot} and for the fair comparison, we explored TW results under varying tuning parameters and the results are shown in Fig.~\ref{fig:TWDWcont}. }
\label{fig:dwTW}
\end{figure}

\begin{figure}
    \centering
    \includegraphics{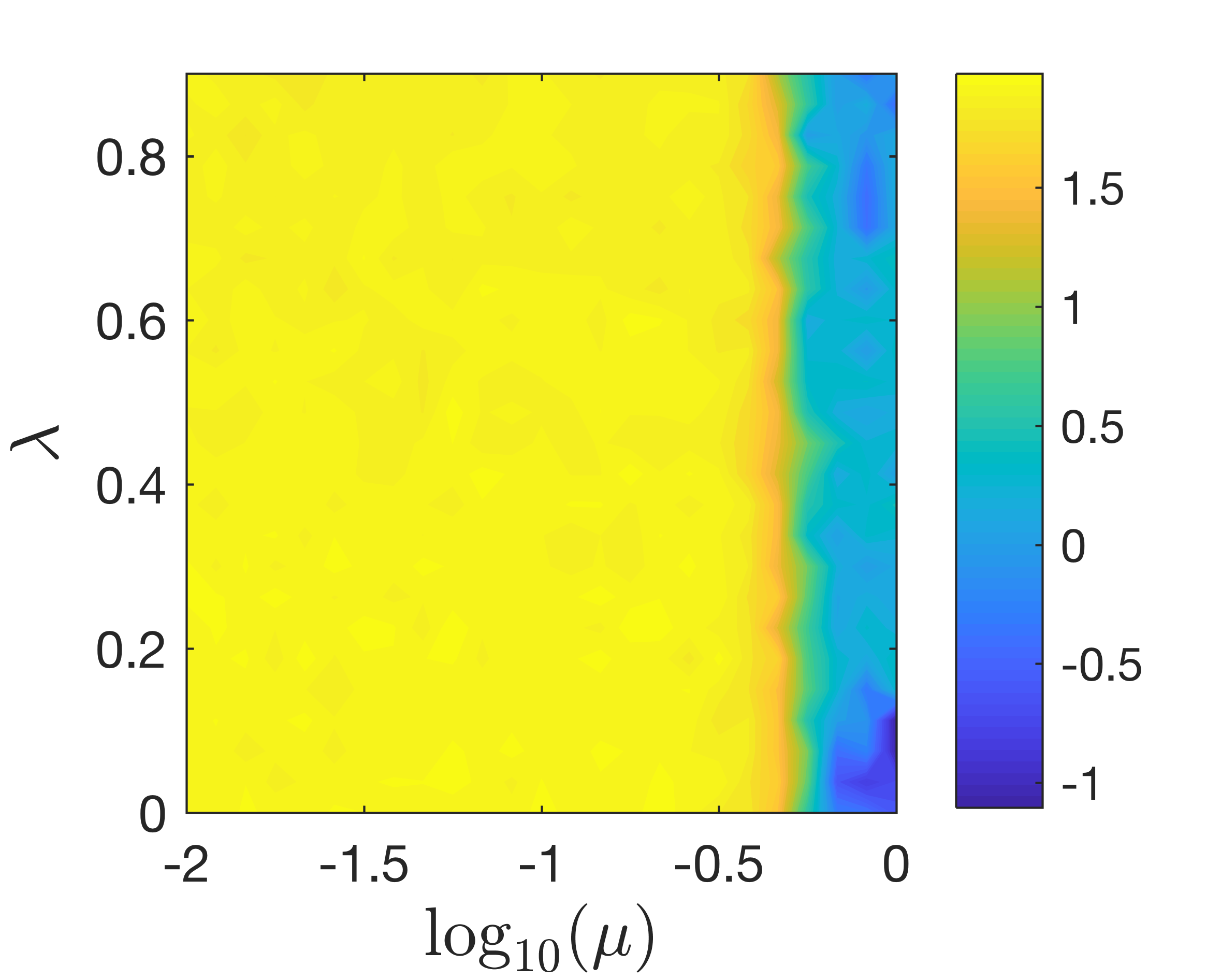}
    \caption{Double Well potential example. Error in recovered solution by TW under different values of $\lambda$ and $\mu$ for the example shown in Fig.(\ref{fig:dwTW}). Although the problem measurements are fixed, TW is also depended in random number generator seed, so, we averaged the results over 100 runs. We see that TW has overall failed in recovering the parameters. Although it has some degree of success in the very narrow lower-right corner with error = 0.1}
    \label{fig:TWDWcont}
\end{figure}

\newpage
\subsection{Lorenz system.}

\begin{figure}
    \centering
    \includegraphics[scale=1]{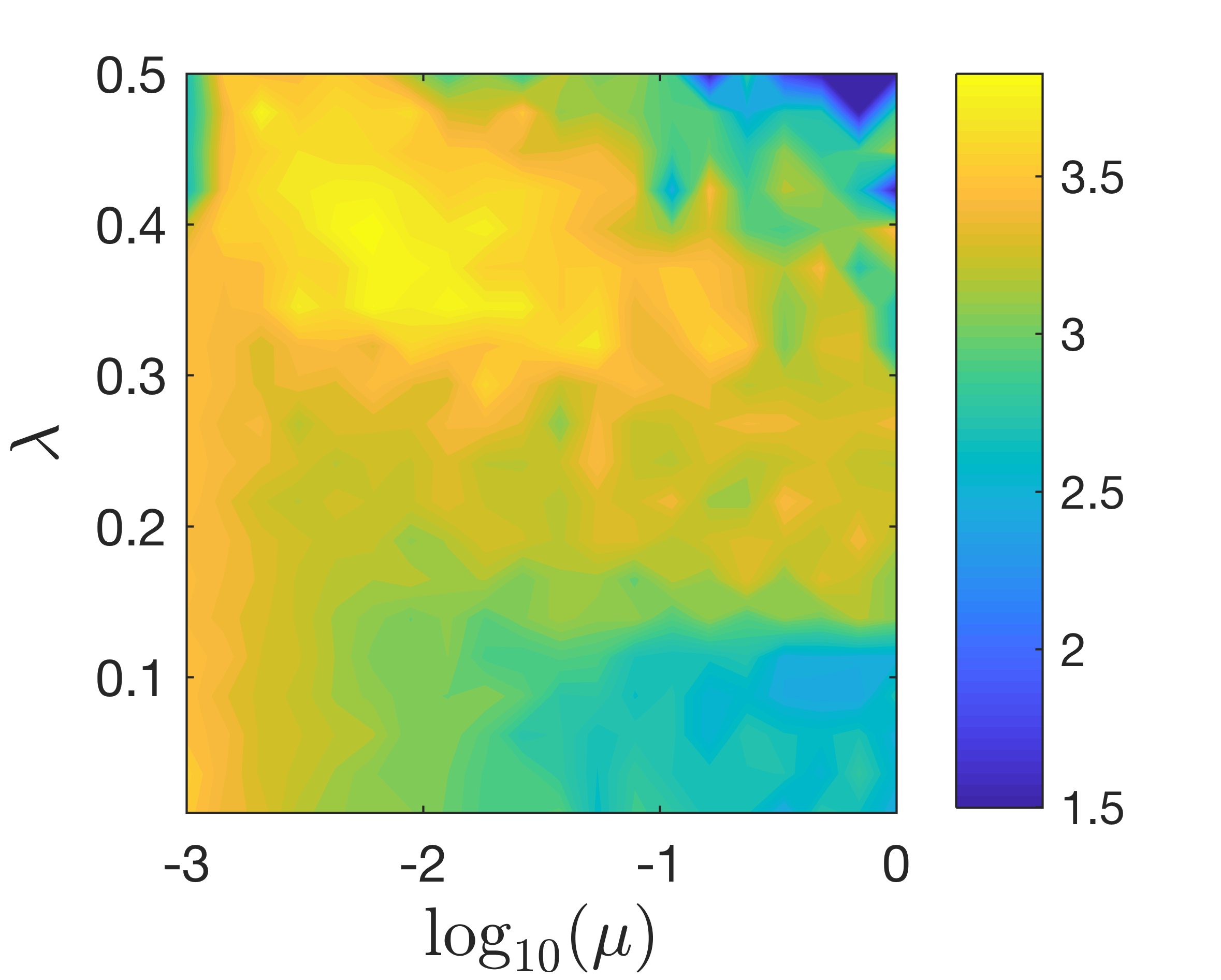}
    \caption{Contour plot of the error in recovered solution of Lorenz system (Fig.(\ref{fig:ER_Lorenz})) by TW method for a grid of $\mu$ and $\lambda$ values and using 2000 measurements, $5^{th}$ order polynomial expansion, low noise with $\epsilon_1 = 10^{-5}$ and no corrupted data. The color bar indicates the value of $\log_{10}(error)$ in the recovered solution, and it shows large error at all levels of tuning parameters.}
    \label{fig:TWcontourplot}
\end{figure}


\begin{figure}
    \centering
    \includegraphics{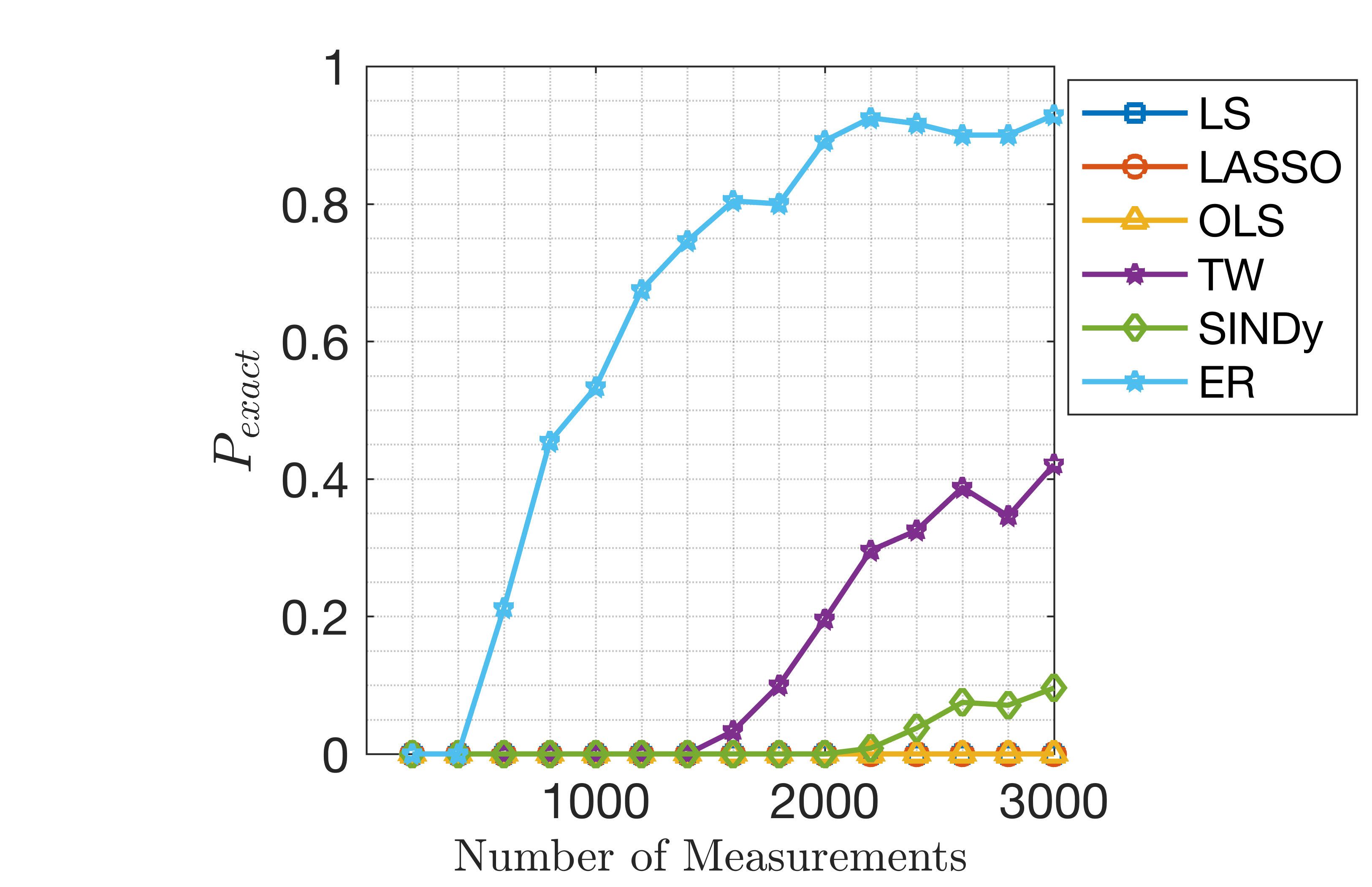}
    \caption{Probability of exact recovery for Lorenz system. For the same results shown in main text Fig.(\ref{fig:ER_Lorenz2}), $P{exact}$ here represent the number of runs in which a method recovered the exact sparse structure over the total number of runs. We see that although TW reached high accuracy at a high number of measurements, its exact recovery probability remains low.}
    \label{fig:exactRecovery}
\end{figure}

\begin{figure}
    \centering
    \includegraphics[scale=1]{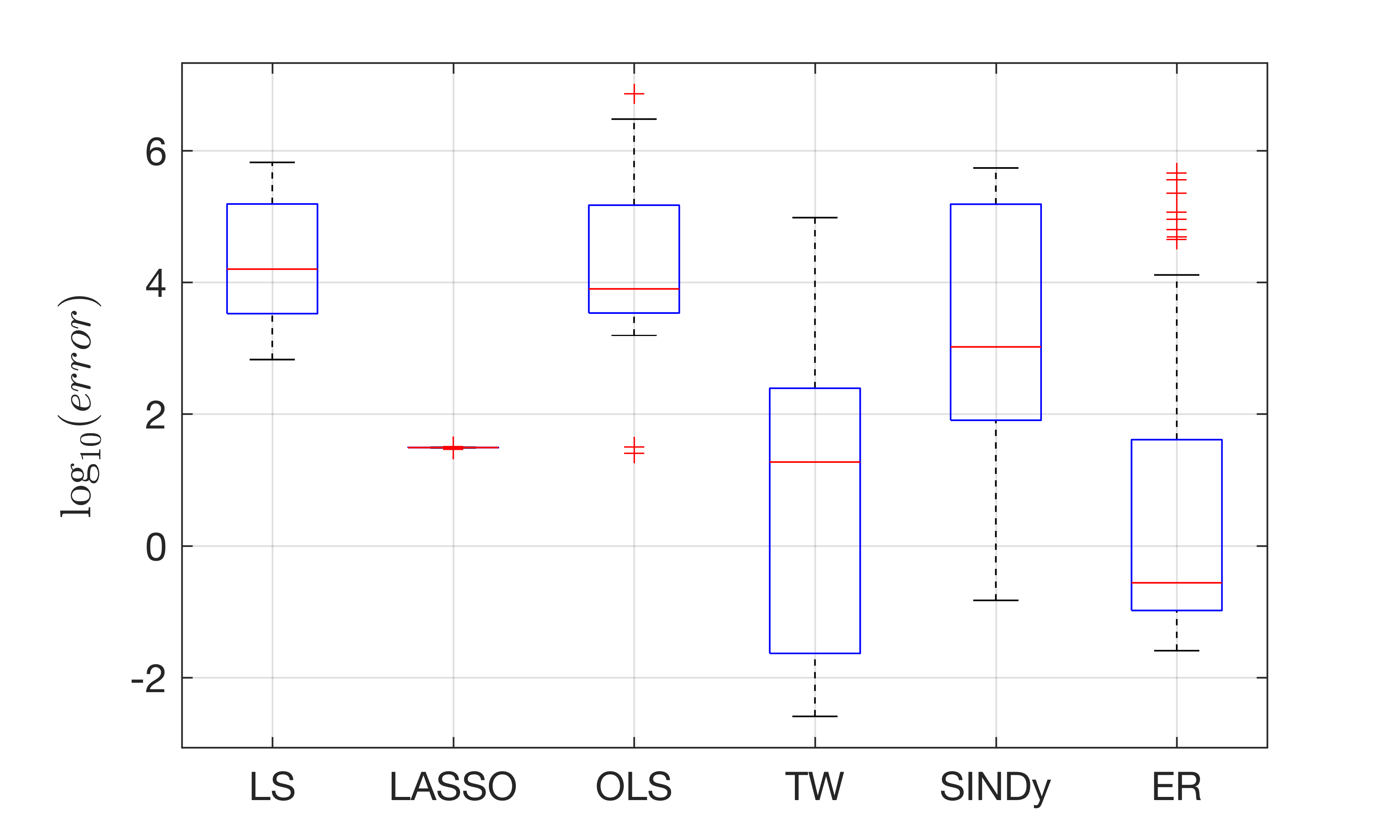}
    \caption{Boxplot for Lorenz. Referes to main text Fig.(\ref{fig:ER_Lorenz2}) at 1500 measurements, this figure shows the results of the all 100 runs.}
    \label{fig:LorenzBoxPlot}
\end{figure}

\newpage
\subsection{Coupled Network of Logistic map}

\begin{figure}
\centering
\includegraphics[scale=1]{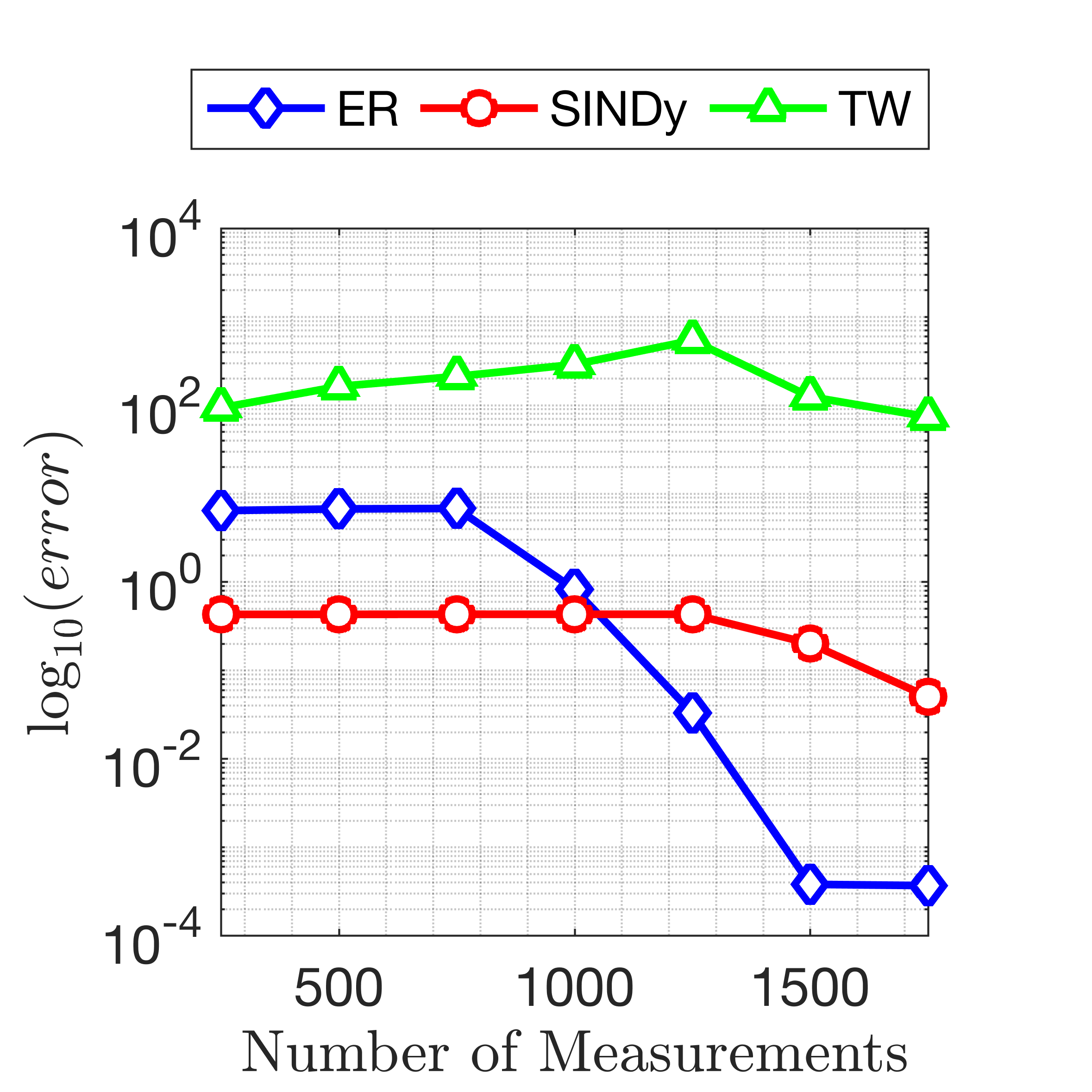}
\caption{Coupled Logistic map example. The error in recovered parameters with noise $\epsilon = 10^{-3}$, second order expansion. As discussed in the Methods section in our main text, we see that TW could not conserve SINDy error level and it diverge to higher error levels until SINDy starts to slightly converge (but still with high error) to the solution with 1500 measurements. While we see that 1500 measurements were enough for the ER to recover the exact sparse structure with high accuracy.}
\label{fig:coupledLogistic}
\end{figure}

Our third example is a network of coupled logistic maps which is a generalization of coupled map lattices \cite{Kaneko1992} and also cellular automata \cite{Kaspar1987}. This scenario of high dimensional and complex systems that have become the thrust of recent analysis including in the synchronization literature \cite{Masoller2001, Anteneodo2006, Jalan2005}. In this example, we assume that not only the governing dynamics are unknown, but so is the structure of the network that moderates the coupling between individual chaotic elements; both of these must be (simultaneously) identified from observed dynamic data alone. In Fig.~\ref{fig:coupledLogistic}, we compare the results of several system identification methods, including the proposed ER approach.  We now offer here a rough description of why this dramatic difference in performance, in the setting particular here of noisy data subject to outliers; a more detailed mathematical analysis will be the subject of our future work.

Consider that each of these other methods we reviewed involves minimizing a functional $J({\mathbf a})$ of the data $a$, and that when ${\mathbf a}$ is subject to noise, that the functionals are each continuous with respect to their argument. We assume that the underlying system is, 
\begin{equation} \label{eq:logisticMap}
f(x) = ax(1-x),
\end{equation}
describing the individual elements as Logistic maps, but,
the coupled network of $N$ such oscillators is of the form,
\begin{equation} \label{eq:logisticMapCoupled}
F(x_i) = f(x_i) + k\sum_{j=1}^{N} A_{ij}\left( f(x_j) - f(x_i)\right)
\end{equation}
\textcolor{black}{\noindent where $i,j = 1,...,N$, $A$ is the adjacency matrix of the coupled network, $k$ is the global coupling strength, and $f(x_i)$ is the image of the point $x_i$ under the logistic map given in Eq.\ref{eq:logisticMap}.}

\begin{figure}
    \centering
    \includegraphics{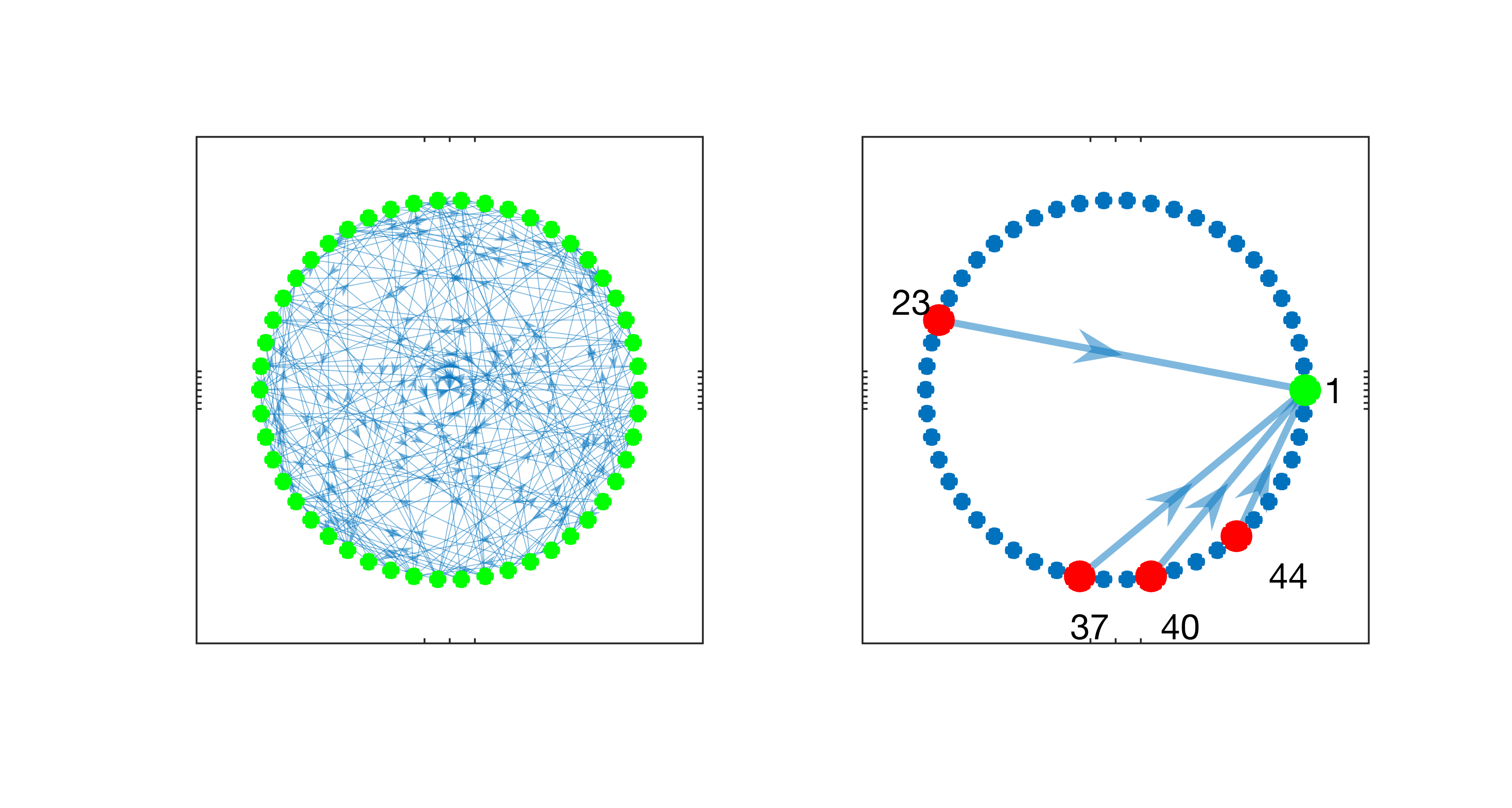}
    \caption{Graph representation of the coupled network of logistic map example. (Left) The 50-nodes network in the directed graph representation. (Right) For a selected node, we see that it is basically influenced by few other nodes.}
    \label{fig:logisticgraph}
\end{figure}

\noindent To present a specific example, let $N=50$, we construct the adjacency matrix $A$ to have simple coupling such that:
\begin{equation} \label{logisticmapcouplingbound}
1 < D_{ii} \leq 4
\end{equation}
where $D$ is the degree matrix of $A$, and the coupling adjacency matrix $A$ constructed randomly such that the above inequality holds. Fig.~(\ref{fig:logisticgraph}) show the graph of the coupled network. Then if we consider only the second order expansion (where the basis matrix $\Phi$ is the second order expansion of the 50 time-series of all nodes) we will have 1326 terms in our expansion matrix. We focus on this example on solving the underdetermined system by considering 2000 measurements as maximum available measurements. So, exclude OLS which only solve overdetermined systems and cannot be investigated at a number of measurements less than 1326, and we exclude LASSO and CS for their high computation complexity. Fig.~\ref{fig:coupledLogistic} shows the error in recovered parameters for this example. For simplicity and the computation complexity, we performed the experiment to find the parameters for one single dimension, and results are averaged over 50 runs. 

\textcolor{black}{This example shows the robustness of ER in recovering the coupling structure in complex coupled networks. The computations complexity in such problem can be highly reduced by considering basic and trivial assumptions. For example, we can consider each node $N_i$ as a default influence source for itself, and then instead of starting the forward step in ER from the empty set, we may initialize the index set with the terms that purely includes $N_i$.}

\begin{figure}
    \centering
    \includegraphics[scale=0.8]{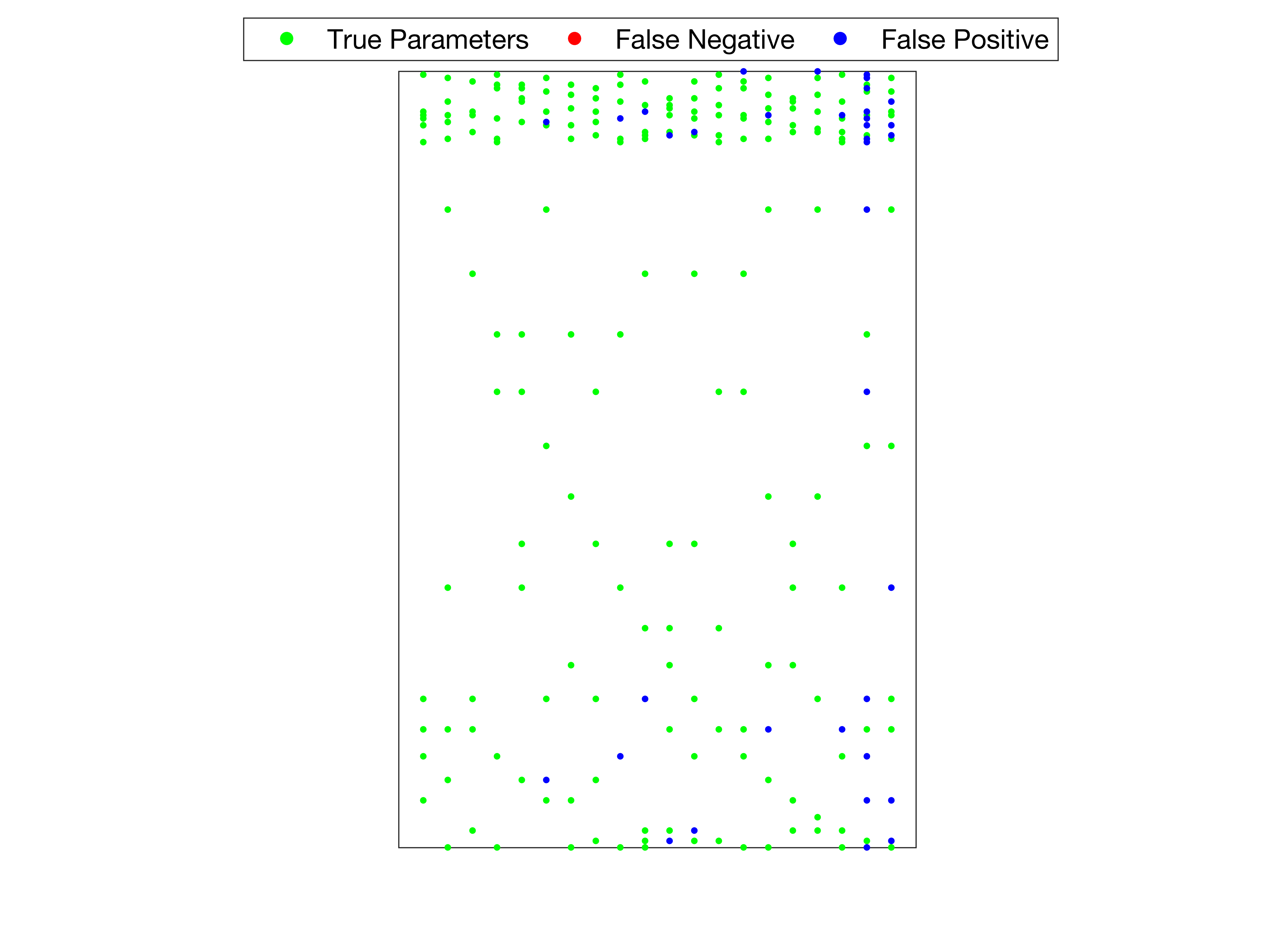}
    \caption{ER solution sparse representation for the coupled Logistic map created by Eqs.~(\ref{eq:logisticMap} - \ref{logisticmapcouplingbound}), with $\epsilon_1 = 0.001$, $\epsilon_2 = 0$, and using 2000 measurements and number of nodes $N=20$. The true solution contained 192 non-zero entries (out of 4620, the total number of parameters) and all of them detected accurately (green dots) with Zero false negative rate, and we see that there was few false positives in ER solution which have 226 total non-zero entries.}
    \label{fig:logsparse_20}
\end{figure}

\textcolor{black}{Fig.(\ref{fig:logsparse_20}) shows the sparse representation of the Logistic map discussed with number of nodes $N = 20$.}

\end{document}